\documentclass[showpacs,preprintnumbers,
superscriptaddress,amsmath]{revtex4}
\usepackage{epsfig}
\usepackage{graphicx}

\usepackage{bm} 

\newcommand{\Diracslash}[1]{#1\llap{/\kern1pt}}
\newcommand{\diracslash}[1]{#1\llap{/\kern0pt}}

\newcommand{\be}{\begin{equation}}
\newcommand{\ee}{\end{equation}}
\newcommand{\bea}{\begin{eqnarray}}
\newcommand{\eea}{\end{eqnarray}}
\newcommand{\ba}[1]{\begin{array}{#1}}
\newcommand{\ea}{\end{array}}

\newcommand{\bbbc}
{{\mathchoice {\setbox0=\hbox{$\displaystyle\rm C$}\hbox{\hbox
to0pt{\kern0.4\wd0\vrule height0.9\ht0\hss}\box0}}
{\setbox0=\hbox{$\textstyle\rm C$}\hbox{\hbox
to0pt{\kern0.4\wd0\vrule height0.9\ht0\hss}\box0}}
{\setbox0=\hbox{$\scriptstyle\rm C$}\hbox{\hbox
to0pt{\kern0.4\wd0\vrule height0.9\ht0\hss}\box0}}
{\setbox0=\hbox{$\scriptscriptstyle\rm C$}\hbox{\hbox
to0pt{\kern0.4\wd0\vrule height0.9\ht0\hss}\box0}}}}

\newcommand{\pslash}{P \mskip -11mu / }

\begin{document}

\title{Are there plasminos in superconductors?} 

\author{Barbara Betz}
\email{betz@th.physik.uni-frankfurt.de}
\affiliation{
Institut f\"ur Theoretische Physik,
Johann Wolfgang Goethe-Universit\"at,
Max-von-Laue-Str.\ 1,
D-60438 Frankfurt, Germany
}

\author{Dirk H.\ Rischke}
\email{drischke@th.physik.uni-frankfurt.de}
\affiliation{
Institut f\"ur Theoretische Physik,
Johann Wolfgang Goethe-Universit\"at,
Max-von-Laue-Str.\ 1,
D-60438 Frankfurt, Germany
}
\affiliation{
Frankfurt Institute for Advanced Studies,
Max-von-Laue-Str.\ 1,
D-60438 Frankfurt, Germany
}

\date{\today} 

\begin{abstract}

Hot and/or dense, normal-conducting systems of 
relativistic fermions 
exhibit a particular collective excitation, 
the so-called plasmino. 
We compute the one-loop self-energy, the
dispersion relation and the spectral density 
for fermions interacting via attractive
boson exchange. It is shown 
that plasminos also exist in superconductors.

\end{abstract}

\pacs{12.38.Mh, 74.62.Yb} 

\maketitle

\section{Introduction}

In relativistic fermionic systems at high temperature 
and/or high density,
there are two types of fermionic excitations. Besides the 
ordinary excitation branches of particles and 
antiparticles, there are
additional collective excitations, the so-called plasmino and 
anti-plasmino \cite{Klimov:1981ka,Weldon:1982bn}. 
These branches have opposite chirality compared to the 
ordinary excitation branches \cite{Weldon:1982bn}.
They coincide with the normal fermionic branches for vanishing 
momenta. For large momenta, they approach the lightcone and 
their spectral strengths vanish exponentially.
The plasmino (anti-plasmino) branch has a minimum (maximum) 
for small, non-zero momenta.
These excitations have been extensively investigated for 
normal-conducting matter
\cite{Pisarski:1993rf,Braaten:1991dd,Blaizot:1996pu,Thoma:1997dk,
Braaten:1990it,Braaten,Schaefer:1998wd,Weldon:1989ys,Pisarski:1989wb,
Baym:1992eu,Peshier:1998dy,Blaizot:1993bb,Peshier:1999dt,Mustafa:2002pb,Wang:2004tg,Kitazawa:2005pp}, for a review see, 
for instance, Ref.\ \cite{Bellac}.

With the exception of heavy-ion collisions, in the laboratory 
it is hard to achieve sufficiently large temperatures 
and densities such that a relativistic description for fermions 
becomes necessary, even if they are as light as electrons. 
On the other hand, there is a plethora of astrophysical
situations where fermions have to be treated relativistically. 
For instance, the core of compact stellar objects 
could be sufficiently dense to consist of deconfined quark 
matter. The quark Fermi
energy is then of the order of $\mu \sim 500$ MeV. Thus, at 
least the light up and down quark flavors have to be 
considered as relativistic particles, since 
$m \sim 5$ MeV $\ll \mu$.
However, quark matter in compact stellar objects, if 
sufficiently cold, is not a normal-conducting system, but
a color superconductor \cite{Review,Rischke:2003mt}. 
In this paper, we
therefore investigate whether the plasminos known from 
normal-conducting systems survive in a superconductor. This is a 
question of general interest, independent of the specific 
nature of the superconductor (i.e., ordinary, color, etc.). 
To our knowledge, this problem has not been considered 
previously, since electrons can to good approximation
be considered as non-relativistic in the condensed-matter 
context, and plasminos
are absent if the temperature is smaller than the mass 
(at least for zero chemical potential \cite{Baym:1992eu}). 
It is also possible to formulate our expectation 
regarding the existence
of plasminos in superconductors: plasminos are 
low-momentum excitations which, for large chemical
potential $\mu \gg p$, are buried deep down in the Fermi sea. 
On the other hand, superconductivity is a Fermi-surface
phenomenon. We thus expect that superconductivity should 
not exert a destructive influence on the presence of
the plasmino excitations. Nevertheless, it requires an 
explicit calculation to prove this, which is
the purpose of the present paper. We shall see 
that our expectations regarding the existence of 
plasminos in superconductors are confirmed.

The outline of this paper is the following. 
In Sec.\ \ref{II}, we consider a normal-conducting system 
consisting of massless fermions interacting via scalar 
and vector boson exchange. For the sake
of simplicity, we restrict our consideration to zero 
temperature. This case has been studied before by Blaizot and
Ollitrault in Ref.\ \cite{Blaizot:1993bb}. We largely 
confirm their results and extend them by computing the spectral
density. We then study a superconducting system in 
Sec.\ \ref{III}. Section \ref{V}
concludes this paper with a summary of our results. 

Our units are $\hbar=c=k_B=1$ and the metric tensor is 
$g^{\mu\nu}={\rm diag}(+,-,-,-)$. Four-vectors are 
denoted by capital letters, $K \equiv (k_0,\vec{k})$. 
Three-vectors have modulus $k\equiv \vert \vec{k} \vert$
and direction $\hat{k}\equiv \vec{k}/k$. Our computations 
are done in the imaginary-time formalism where 
space-time integrals are denoted as
$\int_X \equiv \int^{1/T}_{0} d\tau\int_{V}\,d^3\vec{x}$, while
energy-momentum sums are 
$\int_K \equiv T\sum_{n}\int d^3\vec{k}/(2\pi)^3$
with $n=0,\pm 1\pm 2, ...$ labeling the Matsubara 
frequencies for bosons, $\omega^b_n=2n\pi T$, and
fermions, $\omega^f_n=(2n+1)\pi T$, respectively.

\section{Normal-Conducting Fermions} \label{II}

In this section the dispersion relation and the 
spectral density of massless normal-conducting
fermions is reviewed in the limit $T\rightarrow 0$ 
\cite{Blaizot:1993bb}. We first investigate the case 
where the interaction
between the fermions is mediated by scalar bosons and 
compute the one-loop fermion self-energy. 
We then use the Dyson-Schwinger equation for the fermion propagator
to perform a resummation of the one-loop self-energy to all orders. From the
thus obtained quasiparticle propagator we determine the dispersion relation and
the spectral density. Finally, we conclude this section 
with a discussion of the case where the interaction
is mediated by vector bosons. Note that this treatment is
not fully self-consistent in the sense that we do not use the result for the
resummed quasiparticle propagator to recompute the
one-loop self-energy for an iteration of the above procedure. 
Moreover, we do not solve a Dyson-Schwinger equation for the 
scalar boson propagator, which would lead to the generation of a boson mass $\sim g \mu$.

\subsection{Self-energy}\label{scalar bosons}

The inverse propagator for massless non-interacting 
fermions can be written in the form
\begin{equation}\label{G0inverse}
G_0^{-1}(P) = \pslash+\mu\gamma_0 \equiv \gamma_0 
\sum_{e=\pm} G_{0,e}^{-1}(P)\, \Lambda^e_{\vec p}\;,
\end{equation}
where we introduced the energy projectors
\begin {equation}
\label{Projektoren0}
\Lambda^{\pm}_{\vec p}=\frac{1}{2}(1 \pm\gamma_{0}
\vec\gamma\cdot\hat{p} )\;,
\end{equation}
and the free inverse propagators for 
positive/negative-energy solutions
\begin{equation}
G_{0,\pm}^{-1}(P) \equiv p_0 + \mu \mp p\;.
\end{equation}
Equation (\ref{G0inverse}) can be easily inverted to give
\begin{equation}
G_0(P) = \sum_{e=\pm} G_{0,e}(P)\, \Lambda^e_{\vec{p}} 
\, \gamma_0\;,
\end{equation}
with $G_{0,\pm}(P) = 1/(p_0 + \mu \mp p)$.

For fermions interacting with scalar bosons, the 
interaction part of the Lagrangian
is of Yukawa-type, ${\cal L}_I= g\bar{\psi}\psi\phi$.
The (perturbative) one-loop self-energy has the form
\cite{Kapusta}
\begin {equation}
\label{Sigma}
\Sigma(P) = -g^{2}\, T\sum\limits_{n}\int 
\frac{d^{3}\vec{k}}{(2\pi)^{3}} \,
 	     {\cal D}_{0}(K-P) \, G_{0}(K) \;.
\end {equation}
We assume the boson to be massless as well, with 
propagator ${\cal D}_{0}(Q)=-1/Q^{2}$, thus there is no
tadpole contribution to the fermion self-energy.  
Performing the sum over the Matsubara frequencies one obtains
\begin {equation}
\label{Sigma2}
\Sigma(P)=-g^{2}\int\frac{d^{3}\vec{k}}{(2\pi)^{3}}
\frac{1}{2E_{b}} \sum_{e = \pm}
\Lambda^{e}_{\vec k} \gamma_{0}\left[
\frac{1-N^{e}_{F}(k)+N_{B}(E_b)}{p_0 + \mu -e (k + E_b)}
+\frac{N^{e}_{F}(k)+N_{B}(E_b)}{p_0+\mu - e (k - E_b) }\right]
\;.
\end{equation}
Here, $E_b=\vert \vec{k}-\vec{p} \vert$ is the 
energy of the exchanged boson, 
$N_F^\pm (E)=[e^{(E \mp \mu)/T}+1]^{-1}$ are 
the thermal distribution functions for 
fermions and antifermions, and $N_B (E)=(e^{E/T}-1)^{-1}$ is the 
corresponding one for bosons.

We now project onto the self-energies for 
positive/negative-energy solutions, 
\begin{equation} \label{Sigmapm}
\Sigma_{\pm} (P) \equiv \frac{1}{2} \, 
{\rm Tr} \left[ \Lambda_{\vec{p}}^\pm\, \gamma_0\, \Sigma(P)
\right]\;,
\end{equation}
and perform an analytic continuation, 
$p_0 + \mu \rightarrow \omega + i \eta$,
where $\omega$ is the (real-valued) fermion energy 
relative to the vacuum \cite{Blaizot:1993bb}.

\begin{figure}[h]
 \centering
  \includegraphics[scale = 0.6]{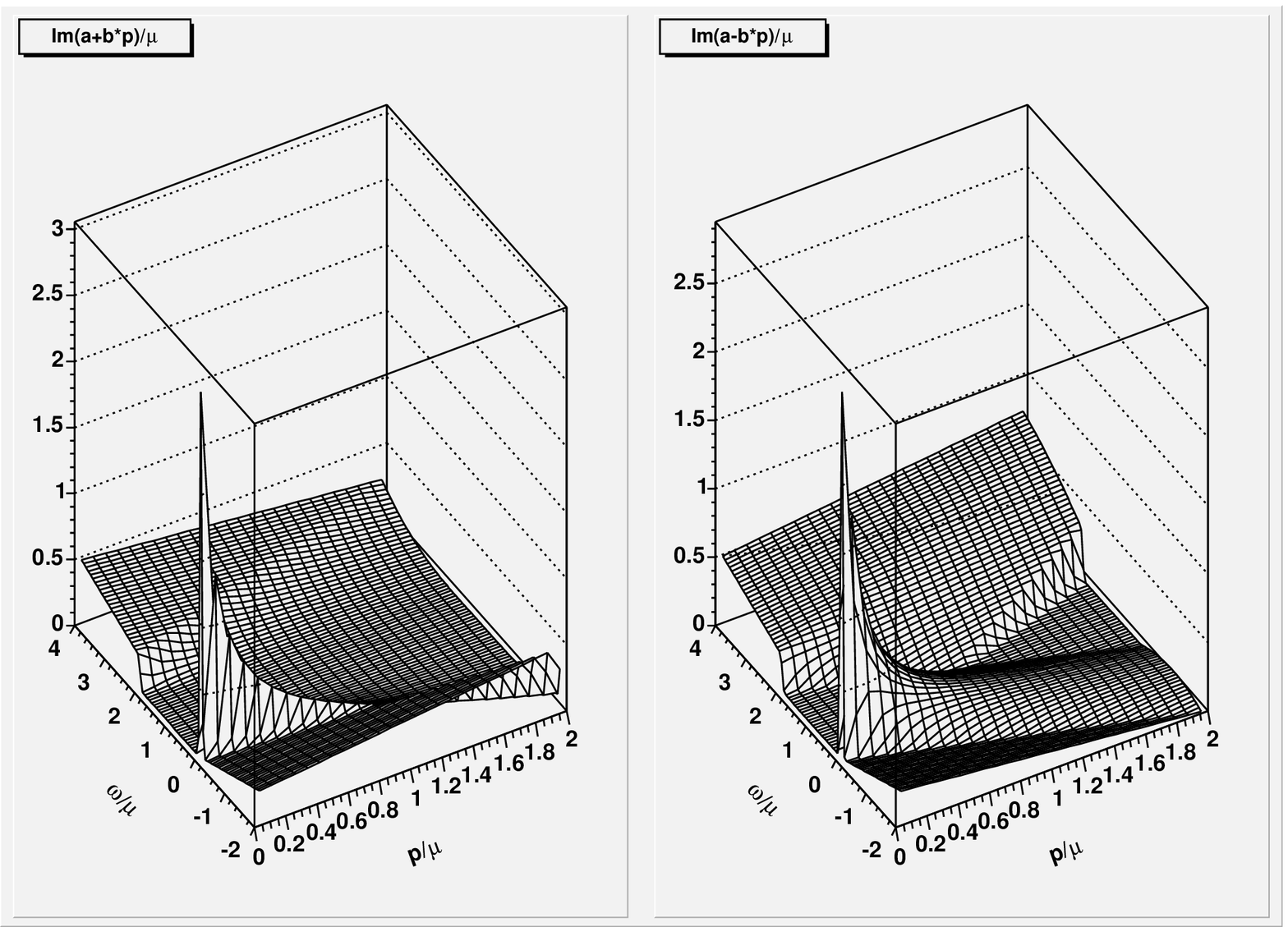}
  \caption{
  The imaginary parts of the self-energy for massless 
fermions (left panel) and antifermions (right panel) at $T=0$
and for an exemplary value of the coupling constant 
$g^2/(4 \pi) =1$.}
  \label{Bild3dIma2}
  \includegraphics[scale = 0.6]{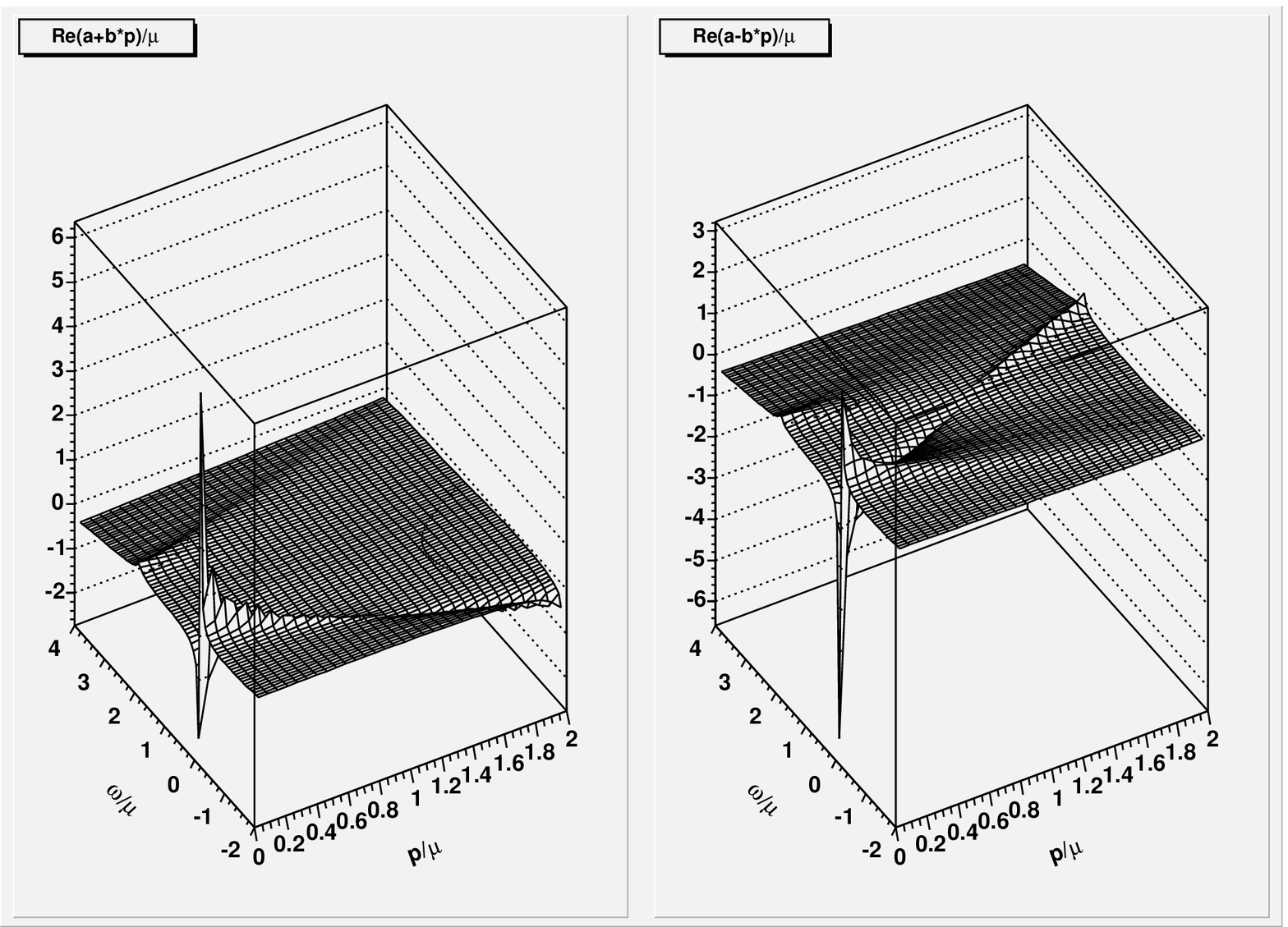}
  \caption{
  The real parts of the self-energy for massless fermions 
(left panel) and antifermions (right panel) at $T=0$ and 
for an exemplary value of the coupling constant 
$g^2/(4 \pi) =1$.}
  \label{Bild3dRea1}
\end{figure}

Using Eq.\ (\ref{Sigmapm}), one readily calculates the 
imaginary part of the self-energy for 
positive/negative-energy solutions,
\begin{equation}
\label{ImSigma}
\hbox{Im}\,\Sigma_\pm(\omega,p)=
\hbox{Im}\,a(\omega, p)\pm
p\, \hbox{Im}\,b(\omega, p)\;.
\end{equation}
Due to rotational invariance, the 
functions $\hbox{Im}\,a(\omega, p)$ and 
$\hbox{Im}\,b(\omega, p)$,
which were first calculated in Ref.\ \cite{Blaizot:1993bb}, 
only depend
on the modulus of the momentum. 
For the sake of completeness we list the results 
in Appendix \ref{AnhangA}.
The real part of the self-energy for 
positive/negative-energy solutions is
\begin{equation}
\label{ReSigma}
\hbox{Re}\,\Sigma_\pm(\omega,p)=
\hbox{Re}\,a(\omega, p)\pm
p\, \hbox{Re}\,b(\omega, p)\;,
\end{equation}
with the functions $\hbox{Re}\,a(\omega, p)$ and 
$\hbox{Re}\,b(\omega, p)$ given in Appendix \ref{AnhangA}.

We show the functions (\ref{ImSigma}) and (\ref{ReSigma}) 
in Figs.\ \ref{Bild3dIma2} and \ref{Bild3dRea1},
respectively. 
The shape of the imaginary and real parts reflect the 
different regions shown in Fig.\ \ref{Energiebereiche}.  
The peak of the imaginary parts in the region of low energies 
and momenta is due to the $1/p$ singularity of the function 
${\rm Im}\,a$, cf.\ Eq.\ (\ref{ErgebnisImaII}). The 
corresponding peak seen in the real parts, however, is 
due to the 
$1/p^2$ singularity of the function ${\rm Re}\,b$, cf.\ 
Eq.\ (\ref{ErgebnisReb}). This is the reason why the 
singularities
appear with the same signs in the imaginary parts, 
but with opposite signs in the real parts of the self-energies 
for particles and antiparticles, cf.\ Eq.\ (\ref{Sigmapm}). 

\subsection{Dispersion relation}

\begin{figure}[ht]
 \centering
  \includegraphics[scale = 0.8]{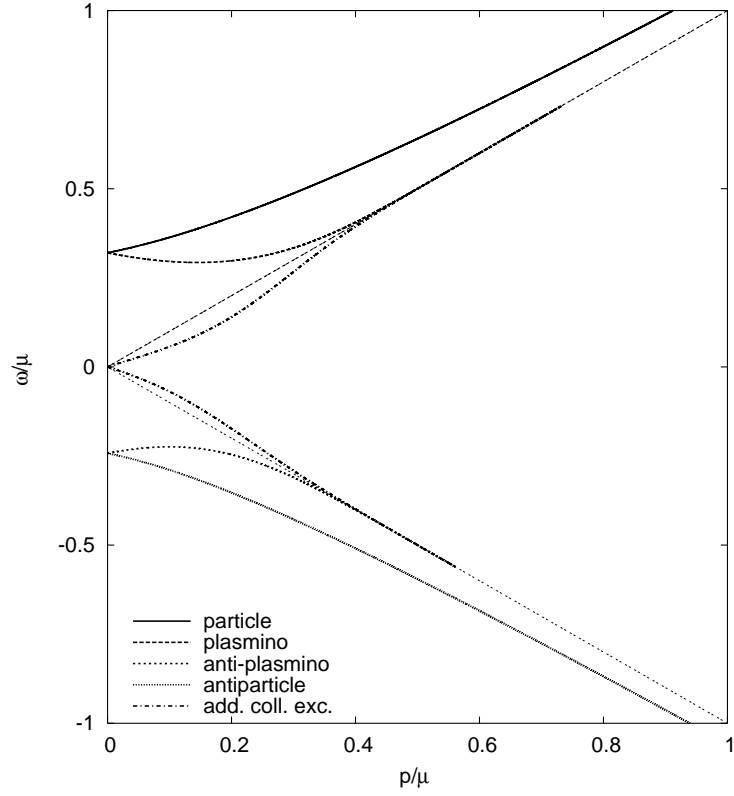}
  \caption{
  Dispersion relations of fermions for $g^2/(4\pi)=1$ at zero
  temperature. The thin lines represent the lightcone.}
  \label{BildDisprel1}
\end{figure}

The full inverse fermion propagator is given by
\begin {equation}
G^{-1}(P)=G_{0}^{-1}(P)+\Sigma(P) \equiv  \gamma_0 
\sum_{e=\pm} G_e^{-1}(P)\, \Lambda^e_{\vec p}\;,
\end{equation}
where the full inverse propagator for 
positive/negative-energy solutions is
\begin{equation}
G_\pm^{-1}(P) \equiv G_{0,\pm}^{-1}(P) + \Sigma_\pm(P)\;.
\end{equation}
The dispersion relations $p_0 + \mu \equiv 
\omega = \omega_\pm^*(p)$ are given by the roots of the real parts of the
inverse propagators,
\begin{equation}
\label{disp}
{\rm Re}\; G_\pm^{-1} (\omega_\pm^*,p) = 0\;.
\end{equation}
We show the solutions of these equations in Fig.\ 
\ref{BildDisprel1}. 
The ordinary particle and antiparticle excitations 
correspond to the uppermost and lowermost curve, respectively. 
Besides
these, for low momenta we find two additional roots of both 
Eqs.\ (\ref{disp}). The two solutions in the time-like
region belong to the plasmino and antiplasmino branches. 
For massless particles right(left)-handedness implies 
positive (negative) helicity, while for massless antiparticles 
right(left)-handedness implies 
negative (positive) helicity. Since plasminos have opposite 
chirality from particles, the plasmino
solution is actually a root of $G_-^{-1}=0$, while the 
antiplasmino 
is a root of $G_+^{-1}=0$. This also holds for the two 
additional solutions in the space-like region. However, since the
imaginary parts for particles and antiparticles are large 
in this region, these excitations are strongly damped and do not
have appreciable spectral weight, cf.\ Fig.\ \ref{BildSpek}. 
The plasmino solutions approach the lightcone and can no longer
be found numerically for large momenta. The size of the 
momentum region where plasminos are found depends on the value 
of the coupling constant. For decreasing coupling constant, 
this region shrinks.

\subsection{Spectral density}

The spectral density is determined from the
relation \cite{Fetter+Walecka} 
\begin {equation}
\label{DefSpek}
\rho_{\pm}(\omega,p)=-\frac{1}{\pi}\, \hbox{Im}\, 
G_{\pm}(\omega,p) \;.
\end {equation}
If the imaginary part of the self-energy vanishes 
(see region IV of Fig.\ \ref{Energiebereiche}), 
the spectral density is proportional to a $\delta$-function 
with support
on the quasiparticle mass shell, $\omega=\omega^*_\pm(p)$,
\begin {equation} \label{deltafunction}
\rho_{\pm}(\omega,p)=\sum_{\omega^*_\pm} \, 
Z_{\pm}(\omega^*_\pm)\, \delta(\omega-\omega^*_\pm)\;,
\;\;\;\; Z_{\pm}(\omega^*_\pm)=\left| \frac{\partial\, 
{\rm Re}\, G_\pm^{-1}(\omega,p)}{\partial \omega} 
\right|_{\omega=\omega^*_\pm}^{-1} \;,
\end {equation}
which corresponds to an infinite lifetime of the associated 
quasiparticle.
This is the case for the plasmino branch and that part of the 
particle excitation branch, that is shown in Fig.\ 
\ref{BildDisprel1}.
(For larger momenta, 
the particle excitation branch enters region 
Ia of Fig.\ \ref{Energiebereiche} where
the imaginary part is non-zero and, consequently, 
the particle excitation becomes unstable.) 
On the other hand, a non-zero imaginary part gives 
rise to a non-zero width of the spectral density,
\begin {equation}
\label{rho-}
\rho_\pm(\omega,p)=\frac{1}{\pi}\, \frac{\hbox{Im}\, 
\Sigma_\pm(\omega, p)}
{\left[G_{0,\pm}^{-1}(\omega, p)+\hbox{Re}\, 
\Sigma_\pm(\omega, p)\right]^2
+\left[\hbox{Im}\, \Sigma_\pm(\omega, p)\right]^2}\;.
\end {equation}
This leads to a finite quasiparticle lifetime which is 
inversely proportional to the width of the spectral 
density around the quasiparticle mass shell. 
This is the case for all other fermionic excitation 
branches which lie outside region IV of Fig.\ 
\ref{Energiebereiche}.

\begin{figure}[ht]
 \centering
  \includegraphics[scale = 0.395]{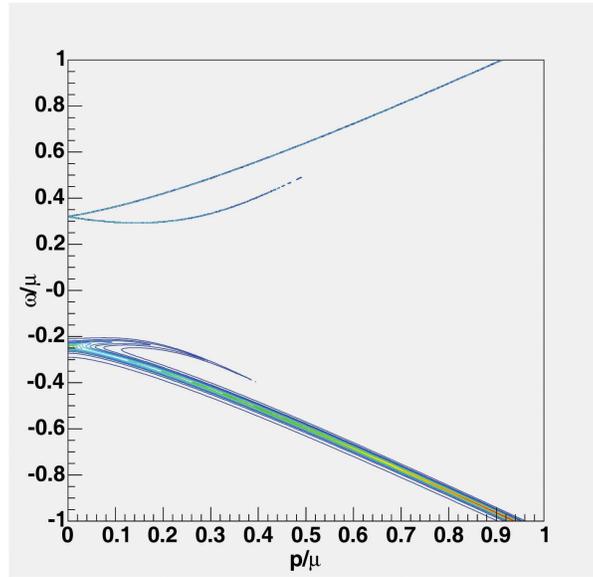}
  \caption{
  Contour plot of the spectral density of fermionic 
  excitations in the limit of vanishing temperature and for 
  $g^2/(4\pi)=1$.}
  \label{BildSpek}
\end{figure}

In Fig.\ \ref{BildSpek} we show the spectral density in a 
contour plot. Comparing this figure to Fig.\ 
\ref{BildDisprel1}, one can easily distinguish the particle, 
plasmino, antiplasmino, and antiparticle branches (from top to
bottom). As mentioned above, the two solutions in the 
space-like region do not have sufficiently large spectral 
weight to
appear in the contour plot. In other words, the width of the 
spectral density at the respective mass shell is large and,
consequently, these excitations are strongly damped. 
Note also that the spectral weight of the two plasmino 
branches rapidly
decreases for larger momenta.

\subsection{Fermions interacting via 
vector bosons}\label{vector bosons}

The self-energy for fermions interacting via 
vector boson exchange is 
\begin{equation}
\Sigma_{\rm v}(P)=-g^{2}T \sum\limits_{n}\int 
\frac{d^{3}\vec{k}}{(2\pi)^{3}}\,
 {\cal D}_{\mu\nu}(K-P)\, \gamma^{\mu}\,
 G_{0}(K)\,\gamma^{\nu}\;.
\end{equation}
For massless vector bosons, this is a gauge-dependent 
expression. One easily convinces oneself that, 
in Feynman gauge for the gauge boson propagator, 
$\Sigma_{\rm v}(P)=2\,\Sigma(P)$. It was shown in Ref.\ 
\cite{Blaizot:1993bb} that the same result holds also in Coulomb
gauge. Finally, in the hard-thermal-loop (HTL) and 
hard-dense-loop (HDL) limit \cite{Weldon:1982bn}, i.e., for 
$T,\mu \gg p_0,p$, the result is gauge-independent, 
and again we have $\Sigma_{\rm v}(P)=2\,\Sigma(P)$, 
provided one takes
the HTL/HDL limit on both sides of this equation. 
Therefore, all these cases are covered by the above discussion 
for scalar boson exchange via rescaling the coupling 
constant, $g \rightarrow g/\sqrt{2}$.

\section{Superconducting Fermions} \label{III}

We now turn to the question whether there are plasminos in 
superconductors. 
Our rationale is the following. We take the superconducting
ground state as a basis for our calculation.
Then, the fermion in the loop in Eq.\ (\ref{Sigma})
is no longer represented by a 
free propagator, $G_0(P)$, but by
a quasiparticle propagator appropriate for 
superconducting systems, ${\cal G}_0$. 
We compute the one-loop self-energy for quasiparticle
excitations in the superconductor.
Afterwards, we resum the one-loop self-energy via
a Dyson-Schwinger equation to obtain the propagator
for quasiparticles and charge-conjugate quasiparticles.
The poles of the propagator determine the
dispersion relation.
Finally, we calculate the spectral density.

Note that this treatment is not fully self-consistent. As in the normal-conducting
case, we do not use the resummed 
quasiparticle propagator to obtain a self-consistent result for the
one-loop self-energy.
Moreover, we do not solve a 
gap equation for the superconducting gap parameter. Rather, we shall use an 
exemplary constant value for the gap function of particles,
$\phi_+(P) = 0.25\, \mu$ and set the gap function 
for antiparticles to zero,
$\phi_-(P) = 0$. The assumption of a constant gap function is justified only for point-like 
four-fermion interactions as in e.g.\ the Nambu--Jona-Lasinio model. For a boson exchange
interaction of non-zero range, the gap function does depend on energy and
momentum. Setting the antiparticle gap to zero is justified only in
weak coupling, because then the
contribution of the antiparticle modes to the gap equation is
negligible \cite{Pisarski:1999tv,Pisarski:1999av}. Since antiparticle
modes live far from the Fermi surface, a non-zero value (of reasonable
magnitude) for the antiparticle gap would only have a negligible influence on
our results for the excitation spectrum and spectral density of antiparticles.

Finally, we also do not solve a Dyson-Schwinger equation for the scalar boson 
propagator which would give rise to a mass $\sim g \mu$.
Nevertheless, we expect
that the complications arising from a fully self-consistent treatment,
including the energy-momentum dependence of the gap function and non-vanishing
antiparticle gaps,
will only lead to quantitative, but not qualitative, changes of our results 
concerning the existence of plasminos in superconductors.

Let us finally note that, in a superconductor, fermions carry a charge with respect 
to a local (gauge) symmetry. This symmetry is spontaneously broken by condensation of
Cooper pairs. An implicit assumption of our approach is that interactions
due to exchange of vector bosons of the gauge interaction 
are negligible as compared to scalar boson exchange, i.e., that the one-loop
fermion self-energy is dominated by the scalar interaction.
If the fermions do not carry any charge, we are strictly speaking not considering
a superconductor but a superfluid.

\subsection{Self-energy}

If fermions interact via scalar boson exchange,
and if we only allow for pairing in the even-parity 
channel, the quasiparticle propagator reads 
\cite{Pisarski:1999av,Fugleberg:2002rk}
\begin{equation}
\label{quasipartprop0}
{\cal G}_0(P)=\sum\limits_{e=\pm}
\frac{p_0-(\mu -e p)}{p_0^2-(\mu-ep)^2-\vert\phi_{e}(P)\vert^2}\,
\Lambda^e_{\vec{p}}\, \gamma_0\; .
\end{equation}
(Note that our ${\cal G}_0$ corresponds to $G^+$ 
in Ref.\ \cite{Pisarski:1999av}.)

\begin{figure}[ht]
 \centering
  \includegraphics[scale = 0.6]{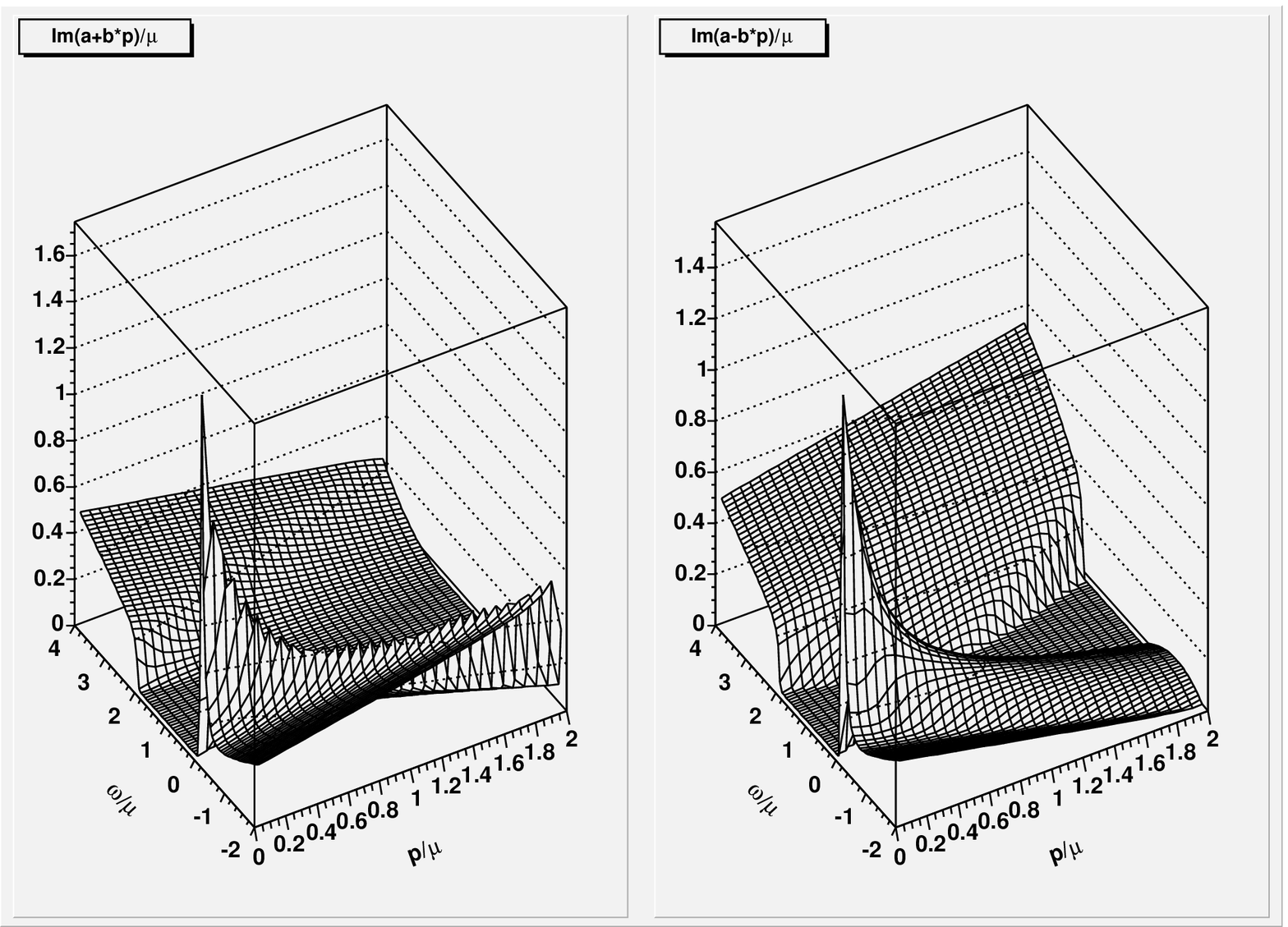}
  \caption{
  The imaginary parts of the self-energy for massless 
  superconducting fermions (left panel)
  and antifermions (right panel), at $T=0$, for 
  $\phi_+ = 0.25\, \mu$ and $\phi_- = 0$.}
  \label{BildT3Ima2}
  \includegraphics[scale = 0.6]{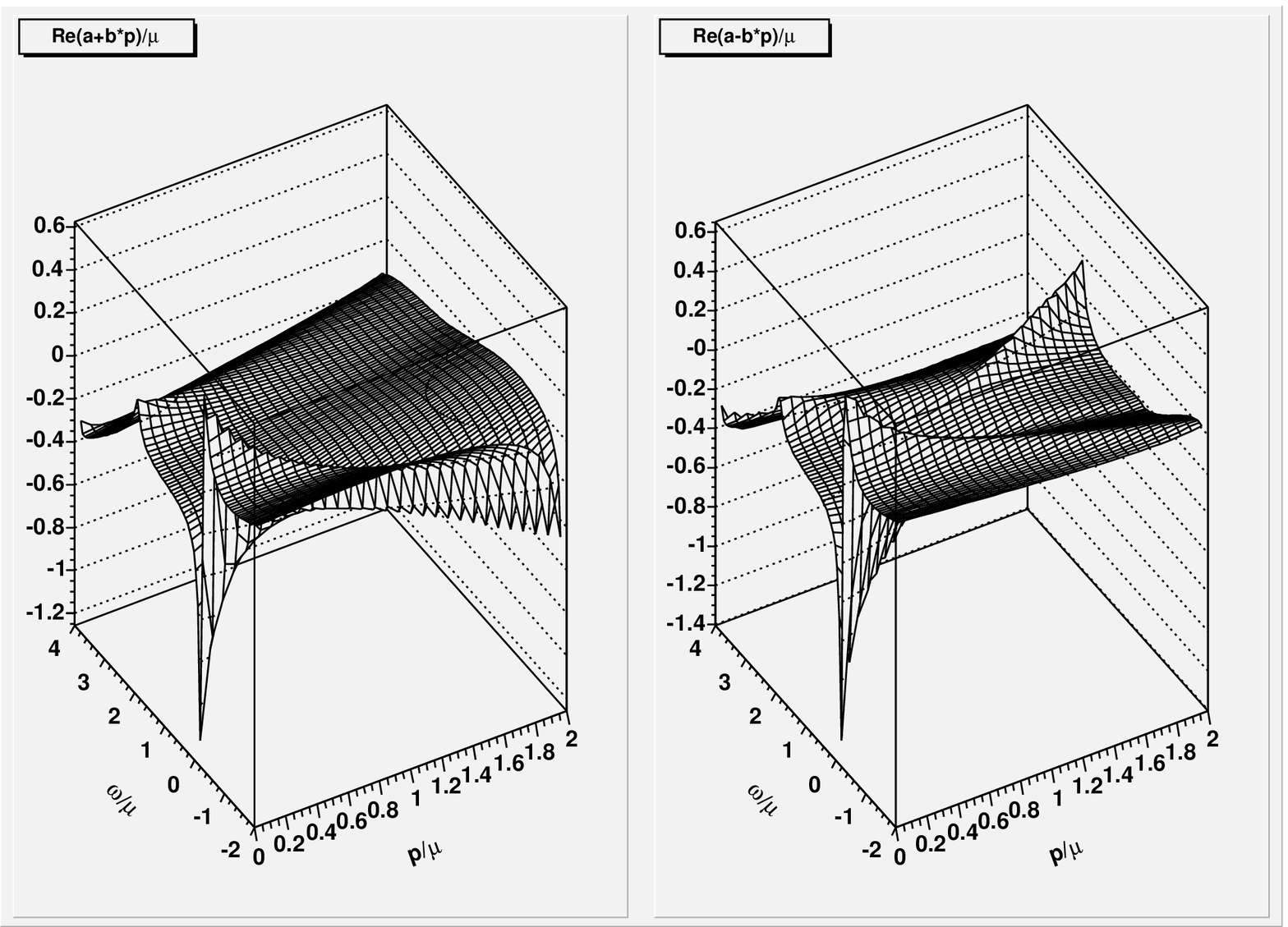}
  \caption{
  The real parts of the self-energy for massless 
  superconducting fermions (left panel)
  and antifermions (right panel), at $T=0$, for 
  $\phi_+ = 0.25\, \mu$ and $\phi_- = 0$.}
  \label{BildT3Rea}
\end{figure}

The analogue of Eq.\ (\ref{Sigma}) for the superconducting 
system now reads
\begin{equation}
\label{SigmaSC}
\Sigma(P) = -g^{2}\, T\sum\limits_{n}\int 
\frac{d^{3}\vec{k}}{(2\pi)^{3}} \,
 	     {\cal D}_{0}(K-P) \, {\cal G}_{0}(K) \;.
\end{equation}
We compute the Matsubara sum using the mixed 
representation for the boson propagator,
\begin {equation}
\label{GemDarBoson}
{\cal D}_0(\tau,\vec q)=\frac{1}{2E_b}\left\{ 
[1+N_B(E_b)]\, e^{-E_b\tau}+ N_B(E_b)\, e^{E_b\tau} \right\}\;,
\end{equation}
as well as the quasifermion propagator,
\begin{equation}
{\cal G}_0(\tau,\vec {k})
=- \sum_{e=\pm} \left[ (1-n_F^e)\, 
\frac{ \epsilon_e - (\mu -ek)}{2\, \epsilon_e}\, 
e^{- \epsilon_e \tau}
+ n_F^e\, \frac{ \epsilon_e + (\mu -ek)}{2\, \epsilon_e}\, 
e^{ \epsilon_e \tau}\right]
\, \Lambda_{\vec k}^e\, \gamma_0\;,
\end{equation}
where $\epsilon_e \equiv \sqrt{(\mu-ek)^2+|\phi_e|^2}$, 
$n_F^e \equiv [e^{\epsilon_e/T}+1]^{-1}$.
At vanishing temperature, the self-energy then becomes
\begin {equation}
\label{SelbstenergieIII}
\Sigma(P)=-g^2\int\frac{d^3\vec{k}}{(2\pi)^3}\frac{1}{4E_b}
\sum_{e=\pm} \Lambda_{\vec{k}}^e\gamma_0 \,\left[
\left( 1 - \frac{\mu - ek}{\epsilon_e} \right) 
\frac{1}{p_0-\epsilon_e-E_b}
+\left( 1 + \frac{\mu - ek}{\epsilon_e} \right) 
\frac{1}{p_0 + \epsilon_e + E_b} \right]\;.
\end{equation}
We now project onto positive/negative-energy states 
according to Eq.\ (\ref{Sigmapm}).
After analytic continuation, the results for the 
imaginary and the real part of the self-energy
can again be written in the form of Eqs.\ (\ref{ImSigma}), 
(\ref{ReSigma}). The functions
${\rm Im}\,a(\omega, p)$, ${\rm Im}\, b(\omega,p)$ are 
available in closed form and are given
in Appendix \ref{AnhangC}. 
However, the real parts ${\rm Re}\,a(\omega, p)$, 
${\rm Re}\, b(\omega,p)$ have to be computed
numerically. The results are shown in Figs.\ 
\ref{BildT3Ima2} and \ref{BildT3Rea}.

Again, the shape of the imaginary and real parts 
reflects the different energy domains in Fig.\ 
\ref{BildGap-Energie}, which
are slightly different from those in Fig.\ 
\ref{Energiebereiche}. Apart from this, the overall shape 
is rather similar 
to that of the corresponding imaginary and real parts 
for normal-conducting fermions, cf.\ Figs.\ \ref{Bild3dIma2} and 
\ref{Bild3dRea1}. Note, though, that the peaks are somewhat 
broader and flatter. This means that the damping of 
quasiparticles is larger than in the normal-conducting case. 

\subsection{Dispersion relation}

We now compute the dispersion relation for superconducting 
fermions. As usual, it is advantageous to distinguish particles
and charge-conjugate particles by introducing the Nambu-Gor'kov 
basis. The propagator for quasiparticles and
charge-conjugate quasiparticles then reads, cf. Eq.\ (136) of 
Ref.\ \cite{Rischke:2003mt}, 
\begin{equation}
\label{quasipartprop}
{\cal G}^{\pm}=\bigg([G^{\pm}_{0}]^{-1}+\Sigma^{\pm}
-\Phi^{\mp}\left\{ [G^{\mp}_{0}]^{-1}+\Sigma^{\mp}
\right\}^{-1}\Phi^{\pm}\bigg)^{-1}\;.
\end{equation}
Here, the inverse free propagator for particles and 
charge-conjugate particles is 
\begin{equation} \label{freeprop}
[G^{\pm}_{0}]^{-1}(P)=\gamma_0 \sum\limits_{e=\pm}[p_0 
\pm (\mu - ep)]\Lambda^{\pm e}_{\vec{p}}\;.
\end{equation}
Note that $G^{+}_{0}$ is identical with $G_0$ of Sec.\ \ref{II}.
Also, if we set $\Sigma^\pm \equiv 0$, ${\cal G}^+$ becomes 
identical with ${\cal G}_0$ in Eq.\ (\ref{quasipartprop0}).
Furthermore, the self-energy for particles, $\Sigma^+$, is 
the one we computed above, cf.\ Eq.\ (\ref{SigmaSC}). 
In the following, we only need its decomposition in terms of 
projections onto positive and negative energies,
\begin{equation} \label{Sigmaplus}
\Sigma^+(P) = \gamma_0 \sum_{e = \pm} \Sigma_e(P) \, 
\Lambda^e_{\vec p}\;.
\end{equation}
The self-energy for charge-conjugate particles, $\Sigma^-$, 
can then be 
computed from $\Sigma^+$ via \cite{Rischke:2003mt} 
\begin{equation} \label{Sigmaminus}
\Sigma^-(P) \equiv C\big[\Sigma^+(-P)\big]^TC^{-1} =
- \gamma_0 \sum_{e=\pm} \Sigma_{e}(-P)\, \Lambda^{-e}_{\vec p}\;,
\end{equation}
where $C=i\gamma^2\gamma_0$ is the charge-conjugation matrix, 
and where we have used the identity
$C \gamma_\mu C^{-1} = - \gamma_\mu^T$. Finally, $\Phi^+$ is 
the order parameter for condensation. If
condensation occurs in the even-parity channel 
\cite{Pisarski:1999av}, 
\begin{equation} \label{Phiplus}
\Phi^+(P)=\sum\limits_{e=\pm}\phi_e(P)\,\Lambda^e_{\vec p}\,
\gamma_5\;.
\end{equation}
The charge-conjugate order parameter $\Phi^-$ is defined as
\begin{equation} \label{Phiminus}
\Phi^- (P) \equiv \gamma_0 \, \left[ \Phi^+ (P)\right]^\dagger 
\, \gamma_0 = - \sum_{e=\pm}
[\phi_e(P)]^*\,\Lambda^{-e}_{\vec p}\,\gamma_5\;.
\end{equation}

The dispersion relation is given by the poles of the propagator 
(\ref{quasipartprop}). 
Inserting Eqs.\ (\ref{freeprop}) -- (\ref{Phiminus}) into 
Eq.\ (\ref{quasipartprop}), we obtain
\begin{equation}
\label{GSpek}
{\cal G}^{\pm}(P)=\left\{[G_0^\mp]^{-1}(P)+\Sigma^\mp(P)\right\}
\sum\limits_{e=\pm}\Lambda^{\mp e}_{\vec p}\bigg[
\bigg(p_0 + \mu-ep+\Sigma_e(P) \bigg) \bigg(p_0 - \mu+ep-
\Sigma_e(-P)\bigg)
-\vert\phi_e(P)\vert^2\bigg]^{-1}\;.
\end{equation}
In order to see which poles belong to solutions for positive 
or negative energy, respectively,
we perform the following projection of the quasiparticle 
propagator (\ref{GSpek}),
\begin{equation}
{\cal G}^\pm_{e}(P) \equiv \frac{1}{2} \, {\rm Tr} \left[ 
{\cal G}^\pm (P)\, \gamma_0 \Lambda_{\vec p}^e\right]\;.
\end{equation}
This gives
\begin{subequations} \label{Gplusminus}
\begin{eqnarray}
{\cal G}^+_+(P) & = &\frac{p_0 - \mu + p - \Sigma_+(-P)}{
[p_0 + \mu - p + \Sigma_+(P)][p_0 - \mu + p - \Sigma_+(-P)]
- |\phi_+(P)|^2}\;, \\
{\cal G}^+_-(P) & = &\frac{1}{p_0 + \mu + p + \Sigma_-(P)}\;,\\
{\cal G}^-_+(P) & = & \frac{1}{p_0 - \mu - p - \Sigma_-(-P)}\;,\\
{\cal G}^-_-(P) & = &\frac{p_0 + \mu - p + \Sigma_+(P)}{
[p_0 + \mu - p + \Sigma_+(P)][p_0 - \mu + p - \Sigma_+(-P)]
- |\phi_+(P)|^2}\;.
\end{eqnarray}
\end{subequations}

Since we have set $\phi_- = 0$, there is a cancellation of 
terms between numerator and
denominator in the propagators 
${\cal G}_-^+$ and ${\cal G}_+^-$. This leads to a reduction 
in the number of poles.

\begin{figure}
\vspace*{-2cm}
\begin{minipage}[ht]{7cm}
  \hspace*{-4.7cm}
  \includegraphics[scale = 0.75]{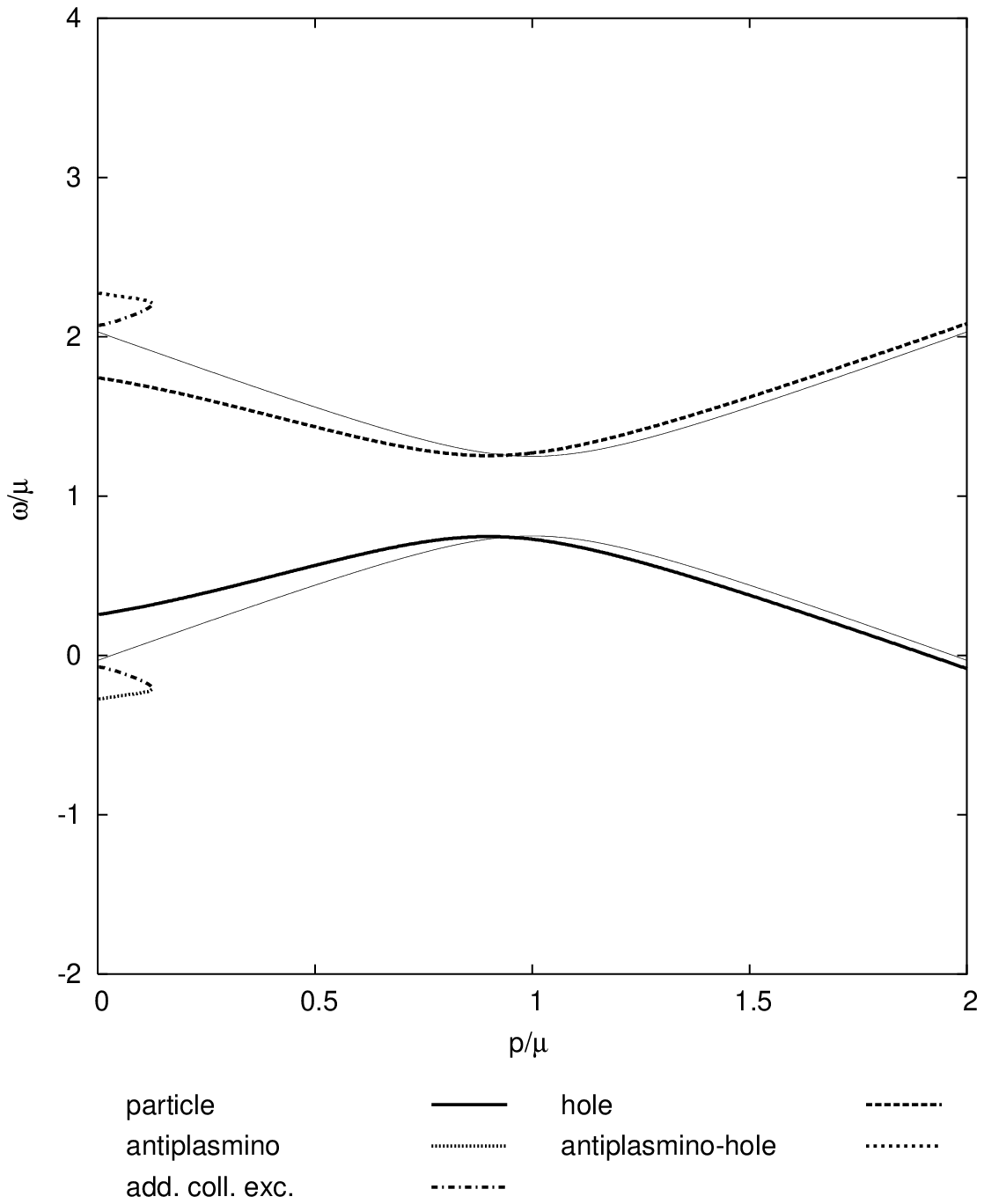}
  \hspace*{-4.2cm}\caption{
  Dispersion relations for superconducting fermions with 
  positive energy for $T=0$, $g^2/(4 \pi) =1$.
  The thin lines are the excitation
  branches for quasiparticles and quasiparticle-holes
  when setting $\Sigma^+ = 0$.}
\label{DisprelSupra1}
\end{minipage} 
\hspace*{1cm}
\begin{minipage}[ht]{7cm}
  \hspace*{-1cm}
  \includegraphics[scale = 0.75]{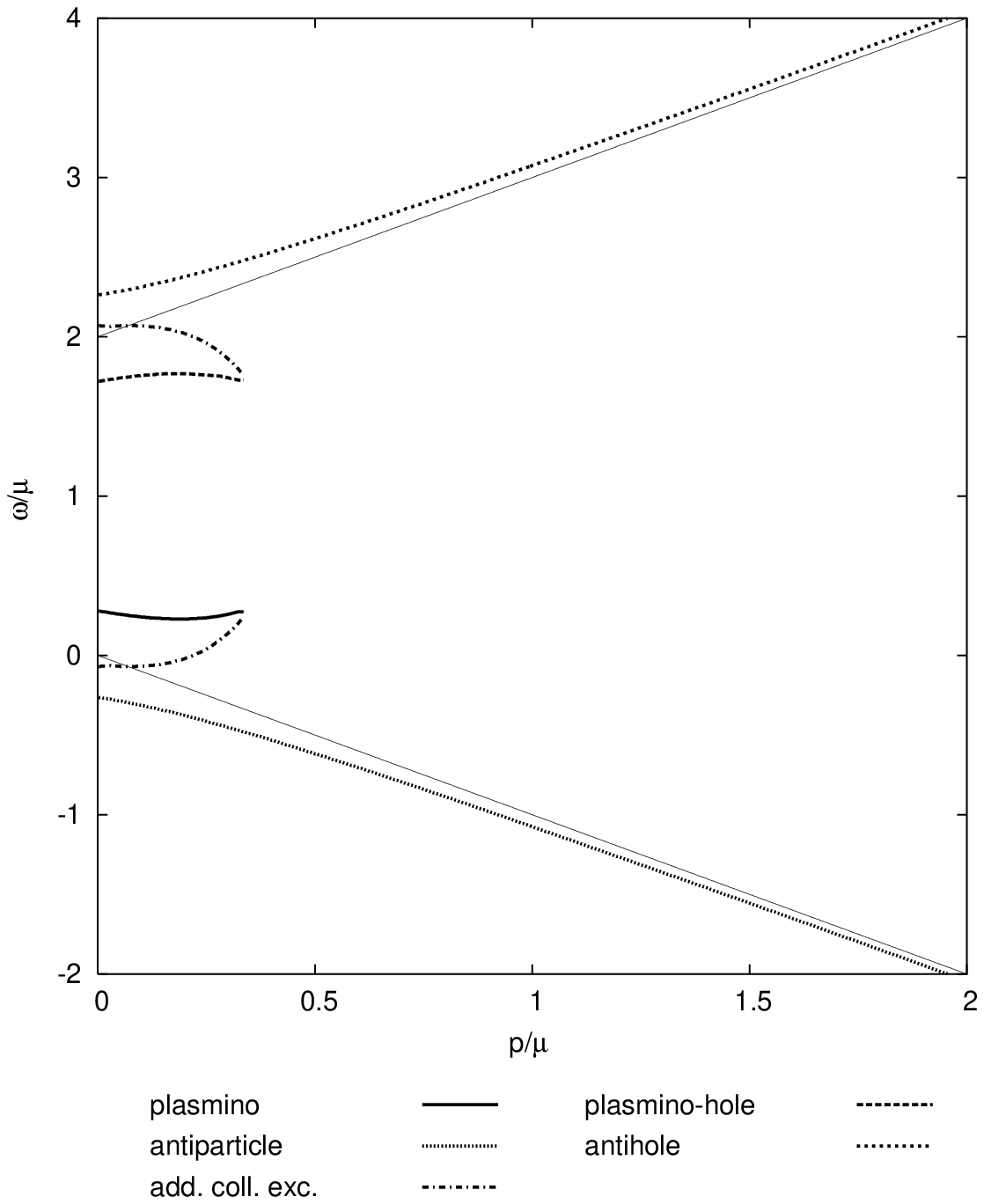}
 \caption{ 
  Dispersion relation for superconducting fermions with 
  negative energy for $T=0$, $g^2/(4 \pi) =1$. 
  The thin lines are the 
  excitation branches of normal antiparticles and 
  antiparticle-holes.}\label{DisprelSupra1b}
\end{minipage}  
\end{figure}

After analytic continuation $p_0 + \mu 
\rightarrow \omega + i \eta$,
we obtain from the poles of ${\cal G}_+^+$ and 
${\cal G}_-^-$ (these propagators have the same poles) 
the dispersion relation for particle, hole, antiplasmino, 
and antiplasmino-hole 
excitations (remember that the latter belong to the 
positive-energy part of the spectrum, and remember also
that charge-conjugate {\em anti-\/}particles correspond to 
positive-energy solutions). 
These dispersion relations are shown in Fig.\ 
\ref{DisprelSupra1} for a value of the 
coupling constant $g^2/(4\pi)=1$. For comparison, 
the thin lines in this figure show
the excitation branches of quasiparticles when setting 
$\Sigma^+ \equiv 0$, i.e.,
the poles of the propagator (\ref{quasipartprop0}) for 
the upper choice of signs.
As seen in the normal-conducting case, Fig.\ 
\ref{BildDisprel1}, there were additional collective
excitations in the space-like domain. We also find the 
corresponding excitations (plus their hole counterparts)
in the superconducting case. These excitations do not 
have an appreciable spectral density (see below), they are 
strongly damped.
The gap in the excitation spectrum is clearly visible. 
However, compared to the gap $2\, \phi_+ = 0.5\, \mu$
in the case $\Sigma^+ =0$ (thin lines), it is shifted with 
respect to the Fermi surface $p=\mu$ to smaller 
values of momentum. This effective shift of the Fermi surface 
is well-known from Fermi-liquid theory
and decreases with the coupling constant $g$.

From the poles of ${\cal G}^+_-$ we obtain the dispersion 
relation of antiparticle and plasmino excitations
(as well as that of the additional excitation), while
the poles of ${\cal G}^-_+$ correspond to plasmino-hole and 
antiparticle-hole excitations (as well as
to the additional excitation in the hole sector). 
These dispersion relations are shown in 
Fig.\ \ref{DisprelSupra1b}. For comparison, we show the 
excitation branches of normal
antiparticles and antiparticle-holes.
As in the normal-conducting case,
plasmino excitations as well as the additional
strongly damped ones exist only in a region of momenta 
smaller than the Fermi momentum.
The size of this region shrinks with decreasing values of 
the coupling constant.

\begin{figure}
\begin{minipage}[t]{7cm}
  \hspace*{-1.7cm}
  \includegraphics[scale = 0.4]{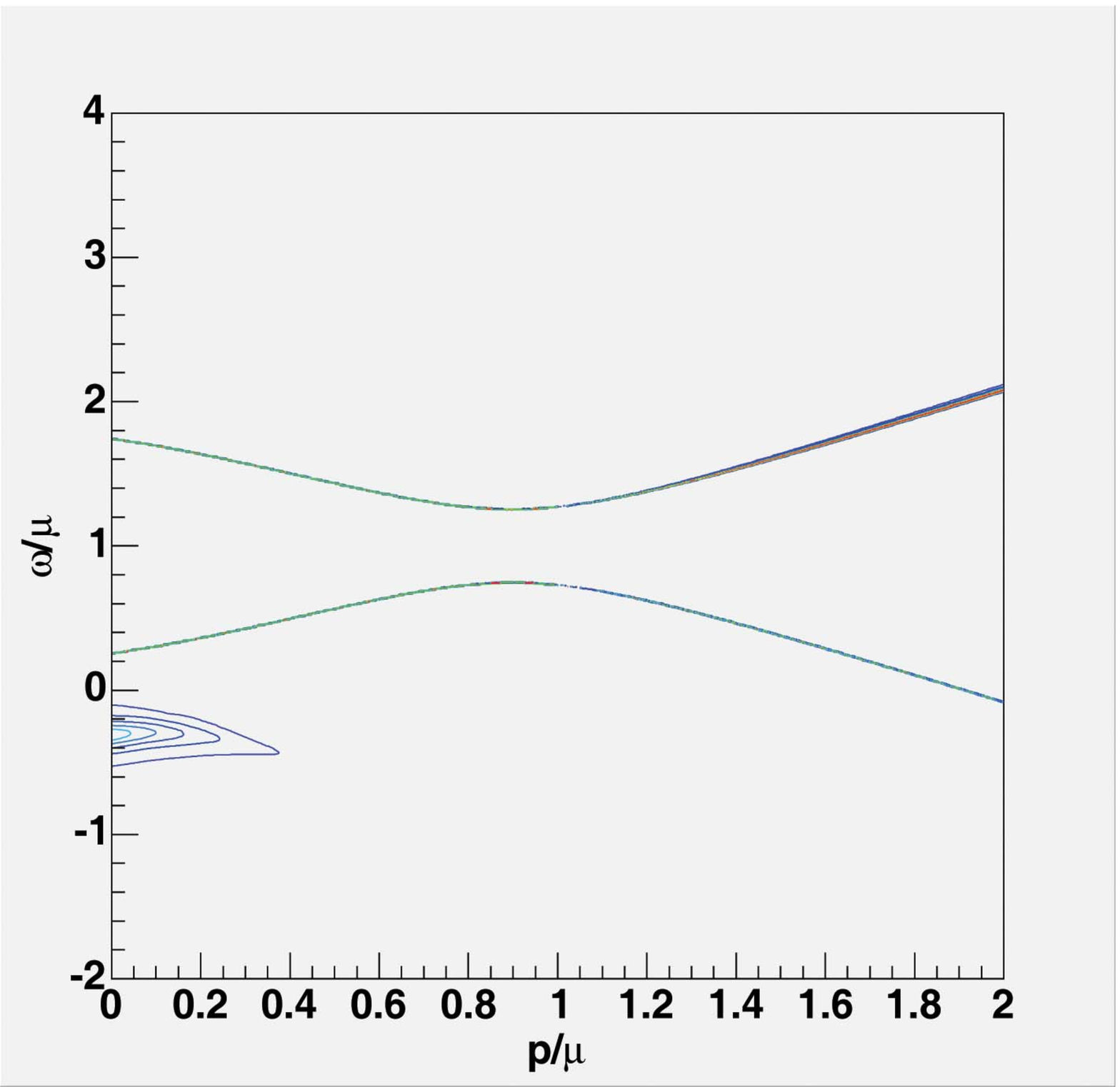}
  \hspace*{-1.7cm}
  \caption{The spectral density $\rho_+^+$ for 
  superconducting fermions, $T=0$,
  $g^2/(4\pi)=1$, $\phi_+=0.25\, \mu$, and $\phi_-=0$. }
  \label{SupraSpek1}
\end{minipage} 
\hspace*{1cm}
\begin{minipage}[t]{7cm}
  \hspace*{-1.7cm}
  \includegraphics[scale = 0.4]{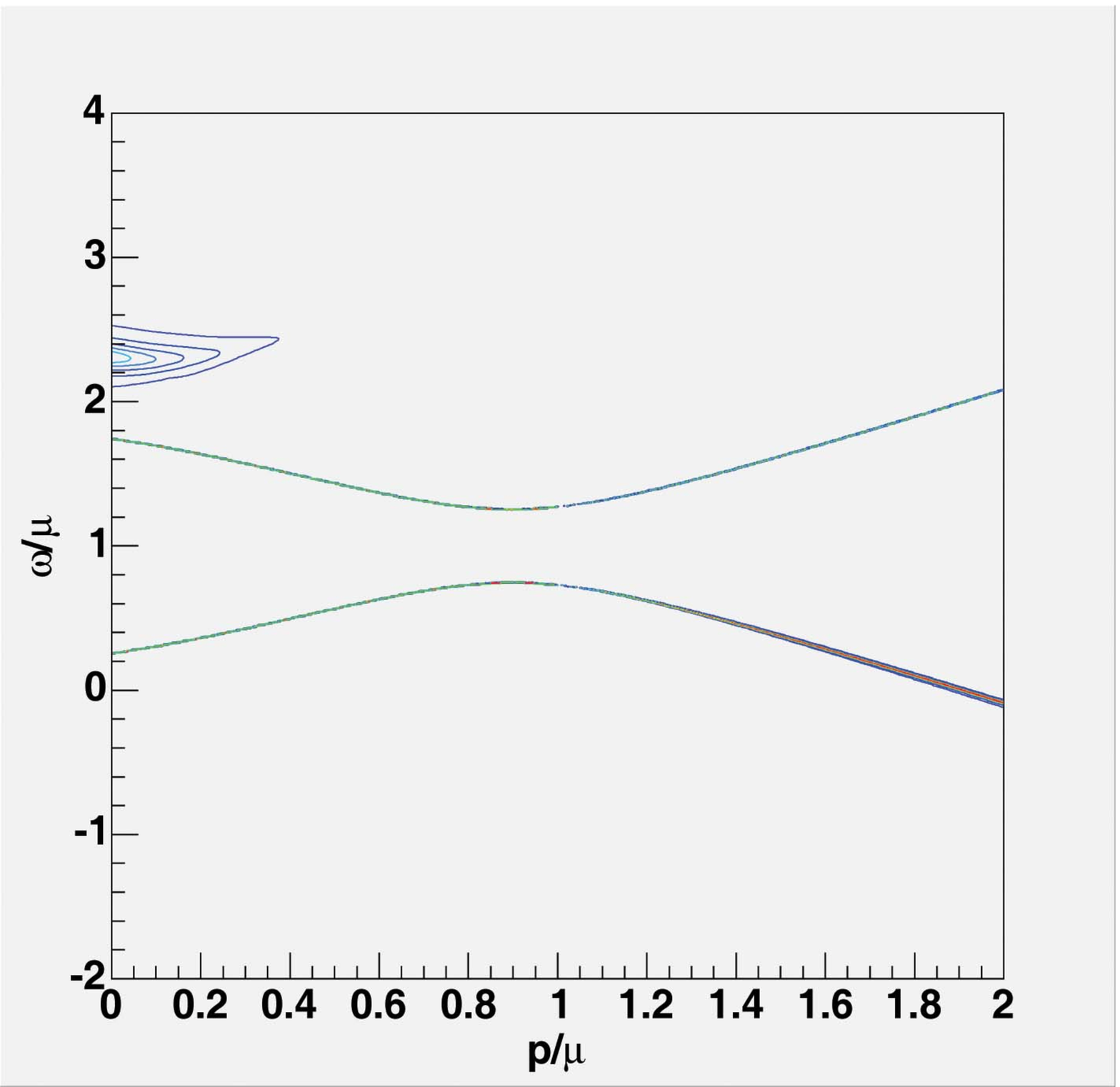}
  \hspace*{-1.7cm}
  \caption{The spectral density $\rho_-^-$ for 
  superconducting fermions, $T=0$,
  $g^2/(4\pi)=1$, $\phi_+=0.25\, \mu$, and $\phi_-=0$. }
  \label{SupraSpek2}
\end{minipage} \\
\begin{minipage}[t]{7cm}
  \hspace*{-1.7cm}
  \includegraphics[scale = 0.4]{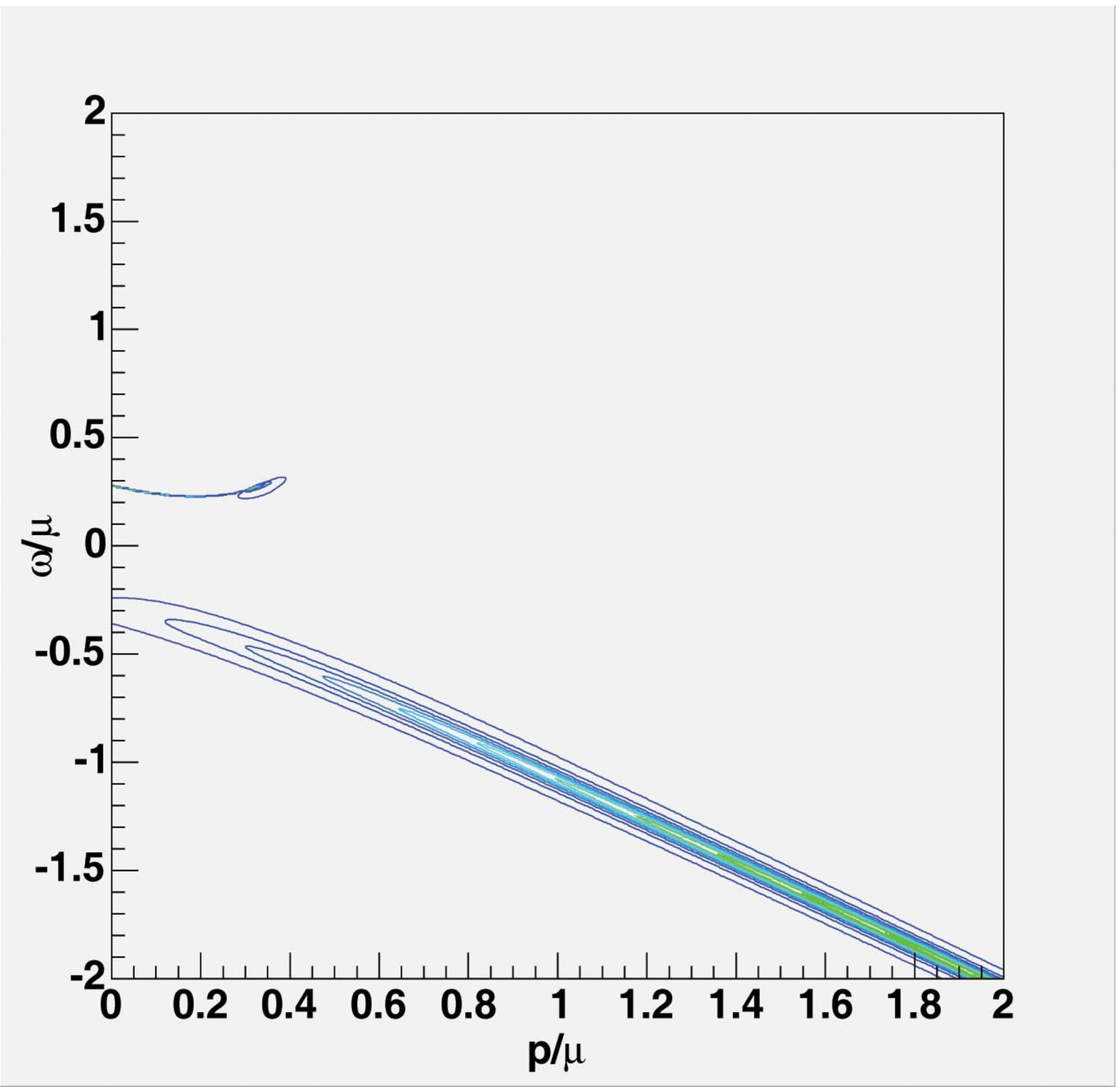}
  \hspace*{-1.7cm}
  \caption{The spectral density $\rho^+_-$ for 
   superconducting fermions, $T=0$,
  $g^2/(4\pi)=1$, $\phi_+=0.25\, \mu$, and $\phi_-=0$. }
  \label{SupraSpek3}
\end{minipage} 
\hspace*{1cm}
\begin{minipage}[t]{7cm}
  \includegraphics[scale = 0.4]{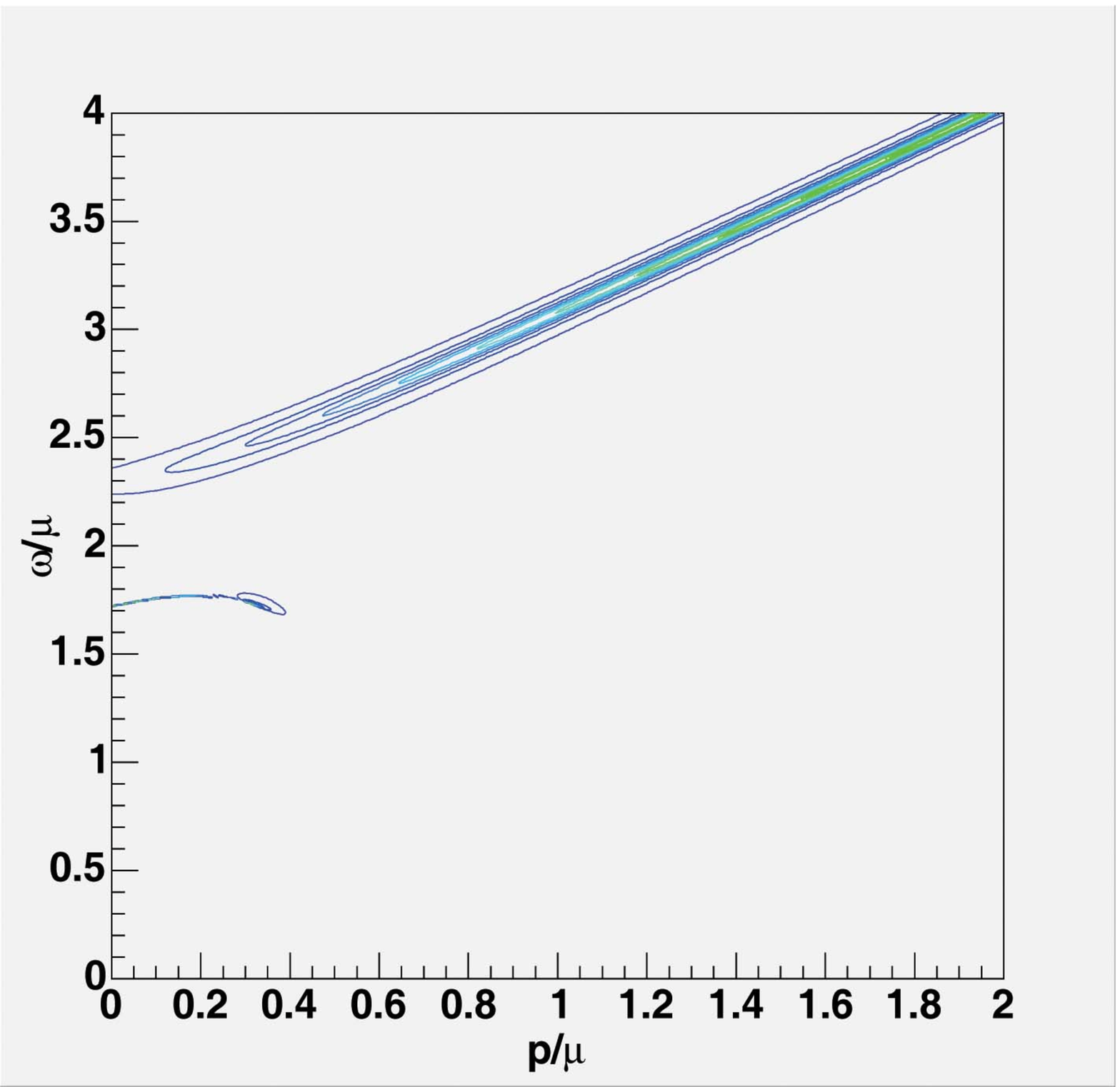}
  \caption{The spectral density $\rho^-_+$ for 
  superconducting fermions, $T=0$,
  $g^2/(4\pi)=1$, $\phi_+=0.25\, \mu$, and $\phi_-=0$.}
  \label{SupraSpek4}
\end{minipage}  
\end{figure}

\subsection{Spectral density}

The spectral densities corresponding to the 
propagators (\ref{Gplusminus})
are computed via
\begin{equation}
\rho^\pm_e(\omega,p) = - \frac{1}{\pi} \, 
{\rm Im}\, {\cal G}^\pm_e(\omega,p)\;,
\end{equation}
$e= \pm$. For the explicit calculation, see 
Appendix \ref{AnhangE}.

In Fig.\  \ref{SupraSpek1} we show a contour plot of 
$\rho^+_+$. For small momenta, the particle and hole branches 
are undamped, 
since in this region the imaginary part of the self-energy 
vanishes. Consequently, the spectral density is
a delta function. For larger momenta, due to a non-zero 
imaginary part of $\Sigma_+$ 
the spectral density broadens on both the particle and the 
hole branch, however,
for the particle branch this broadening is less pronounced 
than for the hole branch. 
One also observes the antiplasmino branch which is strongly 
damped by a non-zero imaginary part of the self-energy.
Although the antiplasmino-hole branch corresponds to a pole 
of ${\cal G}^+_+$, it does not have
appreciable spectral strength in $\rho^+_+$, due to a 
numerical cancellation of terms in the numerator
of the expression for $\rho^+_+$, cf.\ Eq.\ 
(\ref{specdensrho++}). 

In Fig.\ \ref{SupraSpek2} we show 
the spectral density $\rho^-_-$.
The spectral strength on the particle and hole branch is 
comparable to that for $\rho^+_+$.
However, in contrast to that spectral density, now the 
antiplasmino-hole branch is clearly visible
and the antiplasmino branch is suppressed due to a 
numerical cancellation of terms, cf.\ Eq.\ (\ref{specdensrho--}).

In Figs.\ \ref{SupraSpek3} and \ref{SupraSpek4} we 
show the spectral densities $\rho^+_-$ and
$\rho^-_+$, respectively.
We clearly distinguish the plasmino and antiparticle 
branches, as well as the corresponding
hole excitations. The antiparticle and antiparticle-hole 
branches are strongly damped because of the non-vanishing
imaginary part of the self-energy in this region of the 
energy-momentum plane. Note that results similar to ours have been 
obtained in Ref.\ \cite{Nickel:2006mm}, applying the Maximum Entropy Method 
(MEM) to extract the spectral functions from Dyson-Schwinger equations.

\section{Summary} \label{V}

In this work, we have studied the fermionic excitation 
spectrum in normal- and superconducting systems.
For the sake of simplicity, we have taken the fermions 
to be massless and we have considered only the most
simple case of an interaction mediated by massless scalar 
bosons. 
Moreover, we restricted our consideration to zero temperature.
In normal-conducting, relativistic fermionic systems, there
exist additional collective excitations, the so-called 
plasminos. These have
opposite chirality compared to the usual particle excitations.

The goal of the present work was to investigate whether 
such excitations also exist in superconducting systems.
In order to answer this question, we have first computed 
the one-loop fermion self-energy. Using this
self-energy, we have determined the resulting 
dispersion relation from the poles of the full fermion 
propagator.
We do indeed find plasmino excitations in superconducting 
systems.

In addition, we have also identified collective excitations 
which, at least for normal-conducting
systems, lie in the space-like region of the energy-momentum 
plane.
Both the plasmino and these additional collective excitations 
exist only in a region of low energies and momenta.
We then calculated the spectral density and found that the
additional collective excitations are strongly damped and do 
not carry appreciable spectral weight.

\section*{Acknowledgement}

The authors thank Igor Shovkovy and Carsten Greiner 
for discussions and the Center for
Scientific Computing (CSC) of the Johann Wolfgang 
Goethe-University Frankfurt am Main for providing access
to its computing facilities.

\appendix 
\section{Imaginary and real part of the self-energy of 
normal-conducting fermions}
\label{AnhangA}
The imaginary and real part of the self-energy, see 
Eqs.\ (\ref{ImSigma}) and (\ref{ReSigma}), 
respectively, can be computed analytically. 
One distinguishes different regions in the energy-momentum 
plane as shown in Fig.\ \ref{Energiebereiche}.
In these regions, the functions $\hbox{Im}\, a (\omega,p)$ 
and $\hbox{Im}\, b(\omega,p)$ assume the following values, see 
Ref.\ \cite{Blaizot:1993bb}:
\begin{figure}[h]
 \centering
  \includegraphics[scale = 0.75]{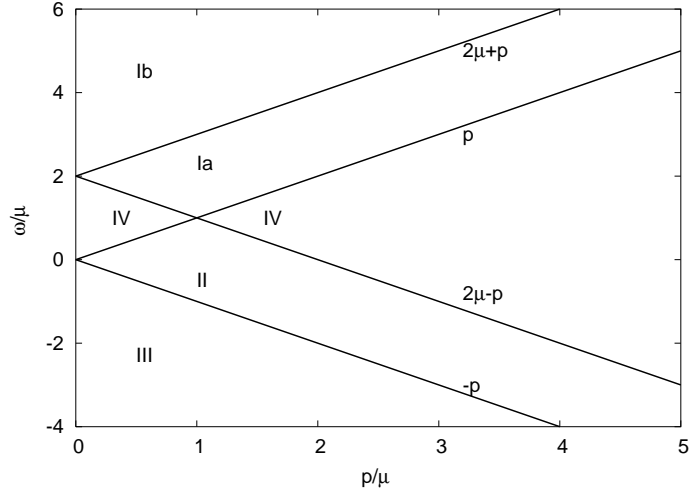}
  \caption{
  The different domains in the energy-momentum plane, 
  which occur in
 the computation of the imaginary part of the self-energy.}
  \label{Energiebereiche}
\end{figure}

\begin{eqnarray}
{\bf Ib:}
\label{ErgebnisImaIb}
\hspace*{0.5cm}\hbox{Im}\,a(\omega,p)&=&\frac{g^2}{32\pi}
\omega\;,
\\
\hbox{Im}\,b(\omega,p)&=&-\frac{g^2}{32\pi}\;,
\vspace*{1.0cm}
\end{eqnarray}\\\vspace*{-1.1cm}
\begin{eqnarray}
{\bf Ia:}
\label{ErgebnisImaIa}
\hspace*{0.5cm}\hbox{Im}\,a(\omega,p)&=&-\frac{g^2}{32\pi}
\frac{1}{4p}(2\mu-\omega-p)(2\mu+\omega+p)\;,
\\
\label{ErgebnisImbIb}
\hbox{Im}\,b(\omega,p)&=&-\frac{g^2}{32\pi}\frac{1}{2p^3}
(2\mu-\omega-p)
\left[\omega^2-p^2-\frac{\omega}{2}(2\mu+\omega+p)\right]\;,
\vspace*{1.0cm}\\
{\bf II:}
\label{ErgebnisImaII}
\hspace*{0.5cm}\hbox{Im}\,a(\omega,p)&=&\frac{g^2}{32\pi}
\frac{1}{4p}(2\mu-\omega-p)(2\mu+\omega+p)\;,
\\
\label{ErgebnisImbII}
\hbox{Im}\,b(\omega,p)&=&\frac{g^2}{32\pi}\frac{1}{2p^3}
(2\mu-\omega-p)
\left[\omega^2-p^2-\frac{\omega}{2}(2\mu+\omega+p)\right]\;,
\vspace*{1.0cm}\\
{\bf III:}
\hspace*{0.5cm}\hbox{Im}\,a(\omega,p)&=&-\frac{g^2}{32\pi}
\omega\;,
\\
\hbox{Im}\,b(\omega,p)&=&\frac{g^2}{32\pi}\;,
\vspace*{1.0cm}\\
{\bf IV:}
\label{ErgebnisImaIV}
\hspace*{0.5cm}\hbox{Im}\,a(\omega,p)&=&
\hbox{Im}\,b(\omega,p)=0 \;.
\end{eqnarray}

The real part of the self-energy is the sum of a vacuum 
and a matter contribution. 
After renormalization one obtains at the renormalization 
scale $\mu$ the result:
\begin{eqnarray}
\hbox{Re}\,a(\omega,\vec p)&=&-\frac{g^2}{32\pi^2 }\Biggl\{\mu+
\frac{(2\mu-\omega-p)(2\mu+\omega+p)}{4p}\ln 
\Bigg\vert\frac{\omega+p}{\omega+p-2\mu}\Bigg\vert
\nonumber\\\vspace*{1.0cm}\nonumber\\
&&\hspace{16.5mm}
-\,\frac{(2\mu-\omega+p)(2\mu+\omega-p)}{4p}\ln 
\Bigg\vert\frac{\omega-p}{\omega-p-2\mu}\Bigg\vert 
+\omega\ln\Bigg\vert \frac{\omega^2-p^2}{\mu^2}\Bigg\vert
\Biggr\}\label{ErgebnisRea}
\\
\hbox{Re}\,b(\omega,\vec p)&=&-\frac{g^2}{16\pi^2}
\Biggl\{\frac{\mu^2}{p^2}-\frac{\omega\mu}{2p^2}
\nonumber\\\vspace*{1.0cm}\nonumber\\
&&\hspace{15mm}
+\frac{(2\mu-\omega-p)}{4p^3}\left[\omega^2-p^2-
\frac{\omega}{2}(2\mu+\omega+p)\right]
\ln \Bigg\vert\frac{\omega+p}{\omega+p-2\mu}\Bigg\vert
\nonumber\\\vspace*{1.0cm}\nonumber\\
&&\hspace{15mm}
-\frac{(2\mu-\omega+p)}{4p^3}\left[\omega^2-p^2-
\frac{\omega}{2}(2\mu+\omega-p)\right]
\ln \Bigg\vert\frac{\omega-p}{\omega-p-2\mu}\Bigg\vert
\nonumber\\\vspace*{1.0cm}\nonumber\\
&&\hspace{15mm}
-\frac{1}{2}\ln\Bigg\vert \frac{\omega^2-p^2}{\mu^2}\Bigg\vert
\Biggr\}\;.  \label{ErgebnisReb}
\end{eqnarray}
Comparing this to Eq.\ (B6) of Ref.\ \cite{Blaizot:1993bb}, 
we obtain an additional term 
$-\omega\mu/(2p^2)$ in Eq.\ (\ref{ErgebnisReb}).

\section{Imaginary part of the self-energy of superconducting 
fermions}
\label{AnhangC}

Superconductivity is a consequence of the formation of Cooper 
pairs at the Fermi surface, i.e., it involves particles, 
but not antiparticles. Therefore, it is permissible to set the 
antiparticle gap function to zero, $\phi_-(P) =0$.
In the following, we abbreviate the particle gap function as 
$\phi_+(P) \equiv \phi$.
The fact that particles have a gap but antiparticles do not 
forces us to compute particle and antiparticle 
contributions to the imaginary part of the self-energy 
(\ref{SelbstenergieIII}) separately.
We therefore split the latter into two parts. Using the 
projection (\ref{Sigmapm}), we obtain
\begin {eqnarray}
\label{ImSelbstenergieIII}
{\rm Im}\, \Sigma_\pm(\omega,p) &= & \sum_{e=\pm} {\rm Im}\, 
\Sigma_\pm^{(e)}(\omega,p)\;, \\
{\rm Im}\, \Sigma_\pm^{(e)}(\omega,p) & = & {\rm Im}\, 
a^{(e)}(\omega,p) \pm p\, {\rm Im}\, b^{(e)}(\omega, p)\;, \\
{\rm Im}\, a^{(e)}(\omega, p) & = & \pi \, g^2 \, 
\int\frac{d^3\vec{k}}{(2\pi)^3}\frac{1}{8E_b}
\,\left[
\left( 1 - \frac{\mu - ek}{\epsilon_e} \right) 
\delta(p_0-\epsilon_e-E_b)
+\left( 1 + \frac{\mu - ek}{\epsilon_e} \right) 
\delta(p_0 + \epsilon_e + E_b) \right]\; , \\
p\, {\rm Im}\, b^{(e)}(\omega, p) & = & - \,e \pi \, 
g^2 \, \int\frac{d^3\vec{k}}{(2\pi)^3}\frac{\hat{p} 
\cdot \hat{k}}{8E_b}
\,\left[\left( 1 - \frac{\mu - ek}{\epsilon_e} \right) 
\delta(p_0-\epsilon_e-E_b)
+\left( 1 + \frac{\mu - ek}{\epsilon_e} \right) 
\delta(p_0 + \epsilon_e + E_b) \right]\;.
\end{eqnarray}

\begin{figure}
 \centering
  \includegraphics[scale = 0.8]{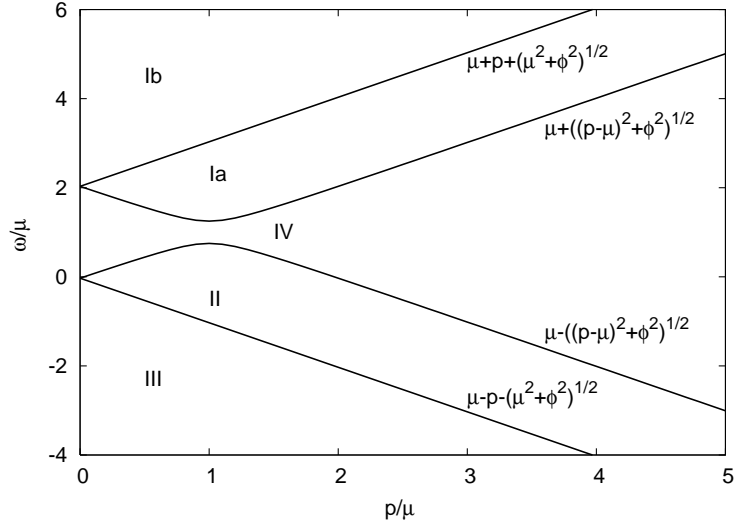}
  \caption{
  The different domains in the energy-momentum plane, 
  which occur in the computation of the positive-energy 
 contribution to the self-energy of superconducting fermions, 
  shown for an exemplary value $\phi=0.25\,\mu$.}
  \label{BildGap-Energie}
\end{figure}

As in the case of normal-conducting fermions,
we encounter different regions in the energy-momentum 
plane when computing the functions 
${\rm Im}\, a^{(e)}(\omega,p)$, ${\rm Im}\, b^{(e)}(\omega,p)$. 
For the negative-energy contribution $(e=-)$
to the self-energy (\ref{SelbstenergieIII}), these regions 
look the same as in Fig.\ \ref{Energiebereiche},
because we have set the antiparticle gap to zero.
However, for the positive-energy contribution $(e=+)$, 
they are of slightly different shape due to the non-zero
value for the particle gap $\phi$. We show these regions 
in Fig.\ \ref{BildGap-Energie}.
 
In the following, we list 
the functions 
${\rm Im}\, a^{(\pm)}(\omega,p)$, ${\rm Im}\, b^{(\pm)}
(\omega,p)$ in the different domains of 
Fig.\ \ref{BildGap-Energie}.

\begin{eqnarray}
\label{Anfang}
\textrm{\bf Ib:}\;\;\;
{\rm Im}\,a^{(+)}&=&\frac{g^2}{32\pi}\omega
\left[1 - \frac{\phi^2}{2 \omega p}
\ln\left(\frac{\omega+p-2\mu}{\omega-p-2\mu}\right)
\right]\; ,\\
{\rm Im}\,b^{(+)}&=&-\frac{g^2}{32\pi}
\left\{ 1 - \frac{\phi^2}{p^2}\left[1 - \frac{\omega - 2 \mu}{2p}
\ln\left(\frac{\omega+p-2\mu}{\omega-p-2\mu}\right)
\right] \right\}\; ,\\
\textrm{\bf Ia:}\;\;\;
{\rm Im}\,a^{(+)}&=&-\frac{g^2}{32\pi}\frac{1}{4p}
\left\{\left(2\mu-\omega-p\right)\left(2\mu+\omega +p\right)
+\phi^2\left[\frac{4\mu}{\omega-p}+\frac{\phi^2}{(\omega-p)^2}
+2\ln\left(\frac{(\omega-p)(\omega+p-2\mu)}{\phi^2}\right)\right]
\right\}\; ,\;\;\;\;\;\;\;\; \\
{\rm Im}\,b^{(+)}&=&-\frac{g^2}{32\pi}\frac{1}{2p^3}
\left\{\left(2\mu-\omega-p\right)\left[
\omega^2-p^2-\frac{\omega}{2}\left(2\mu+\omega+p\right)
\right]\right.
\nonumber\\
&&\hspace*{2cm}
+\frac{\phi^2}{\omega-p}\left[-2\mu p+\frac{\phi^2}{2}
\frac{\omega-2p}{\omega-p}\right]
\left.+\phi^2\left(\omega-2\mu\right)\ln\left(
\frac{(\omega-p)(\omega+p-2\mu)}{\phi^2}\right)
\right\}\; ,\\
\textrm{\bf II:}\;\;\;
{\rm Im}\,a^{(+)}&=&\frac{g^2}{32\pi}\frac{1}{4p}
\left\{\left(2\mu-\omega-p\right)\left(2\mu+\omega +p\right)
+\phi^2\left[\frac{4\mu}{\omega-p}+\frac{\phi^2}{(\omega-p)^2}
+2\ln\left(\frac{(p-\omega)(2\mu-\omega-p)}{\phi^2}\right)\right]
\right\}\; , \;\;\;\;\;\; \\
{\rm Im}\,b^{(+)}&=&\frac{g^2}{32\pi }\frac{1}{2p^3}
\left\{\left(2\mu-\omega-p\right)\left[\omega^2-p^2-
\frac{\omega}{2}\left(2\mu+\omega+p\right)\right]\right.
\nonumber\\
&&\hspace*{1.6cm}
+\frac{\phi^2}{\omega-p}\left[-2\mu p+\frac{\phi^2}{2}
\frac{\omega-2p}{\omega-p}\right]
\left.+\phi^2\left(\omega-2\mu\right)\ln\left(
\frac{(p-\omega)(2\mu-\omega-p)}{\phi^2}\right)
\right\}\; , \;\;\;\;\;\;\\
\textrm{\bf III:}\;\;\;
{\rm Im}\,a^{(+)}&=&-\frac{g^2}{32\pi }\phi^2
\left[\frac{2\mu}{p^2-\omega^2} - 
\phi^2\frac{\omega}{(p^2-\omega^2)^2} 
- \frac{1}{2p}\ln\left(\frac{\omega-p}{\omega+p}\right)
\right]\; , \\
\label{Ende}
{\rm Im}\,b^{(+)}&=&\frac{g^2 \phi^2}{32\pi p^2}
\left[1 - \frac{2\mu\omega}{\omega^2-p^2} - 
\frac{\phi^2 p^2}{(\omega^2-p^2)^2} 
+\frac{2\mu-\omega}{2p}\ln\left(\frac{\omega+p}{\omega-p}\right)
\right]\;,\;\;\;\;\;\;\\
\textrm{\bf IV:}\;\;\;
{\rm Im}\,a^{(+)}&=&{\rm Im}\,b^{(+)}=0\;.
\end{eqnarray}
\\
The negative-energy contribution to the self-energy 
vanishes except
in region III of Fig.\ \ref{Energiebereiche}, where we find
\begin{eqnarray}
\textrm{${\bf III:}$}\,&&
\textrm{Im}\,a^{(-)}=-\frac{g^2}{32\pi}\omega
\\&&\textrm{Im}\,b^{(-)}=\frac{g^2}{32\pi}\;.
\end{eqnarray}

\section{Spectral density of superconducting fermions}
\label{AnhangE}
The spectral densities for superconducting fermions are
proportional to $\delta$-functions in the region of the 
energy-momentum plane, where
the imaginary part of the self-energy is zero. In this case,
they can be computed from an equation analogous to Eq.\ 
(\ref{deltafunction}).
On the other hand, if the imaginary part is non-zero, 
they can be computed via
\begin{eqnarray}
\hspace*{-5.5ex}\label{specdensrho++}
\rho^{+}_{+}(\omega,p)&=&\frac{1}{\pi}\frac{1}{A^2+B^2}
\Big\{\big[\omega-2\mu+p-{\rm Re}\Sigma_+(\mu-\omega,-\vec p\,)
\big]\,B+{\rm Im}\Sigma_+(\mu-\omega,-\vec{p}\,)\,A\Big\}\, ,
\\
\nonumber\\
\rho^{+}_{-}(\omega,p)&=&\frac{1}{\pi}
\frac{{\rm Im}\Sigma_-(\omega-\mu,\vec{p}\,)}
{\left[\omega+p+{\rm Re}\Sigma_-(\omega-\mu,\vec{p}\,)\right]^2+
\left[{\rm Im}\Sigma_-(\omega-\mu,\vec{p}\,)\right]^2}\, ,
\\
\nonumber\\
\rho^{-}_{+}(\omega,p)&=-&\frac{1}{\pi}
\frac{{\rm Im}\Sigma_-(\mu-\omega,-\vec{p}\,)}
{\left[\omega-2\mu-p-{\rm Re}\Sigma_-(\mu-\omega,-\vec{p}\,)
\right]^2+\left[{\rm Im}\Sigma_-(\mu-\omega,-\vec{p}\,)
\right]^2}\, ,
\\
\nonumber\\\label{specdensrho--}
\rho^{-}_{-}(\omega,p)&=&\frac{1}{\pi}\frac{1}{A^2+B^2}
\Big\{\big[\omega-p+{\rm Re}\Sigma_+(\omega-\mu,\vec p\,)
\big]\,B-{\rm Im}\Sigma_+(\omega-\mu,\vec p\,)\,A\Big\}\, ,
\end{eqnarray}
where we have abbreviated
\begin{eqnarray}
\hspace*{-9.5ex}
A&=&\left(\omega-\mu\right)^2-\big(\mu- p\big)^2+
\big(\omega-2\mu+ p\big){\rm Re}\Sigma_+(\omega-\mu, \vec p\,)
+\big(p-\omega\big){\rm Re}\Sigma_{+}(\mu-\omega,- \vec p\,)
\nonumber\\
&&-{\rm Re}\Sigma_+(\omega-\mu,\vec p\,){\rm Re}
\Sigma_+(\mu-\omega,- \vec p\,)
+{\rm Im}\Sigma_+(\omega-\mu,\vec p\,){\rm Im}
\Sigma_+(\mu-\omega,- \vec p\,)
- \phi^2\, ,\;\;\;\;\;
\\\nonumber\\
B&=&\big(\omega-2\mu+ p\big){\rm Im}
\Sigma_{+}(\omega-\mu, \vec p\,)
+\big(p-\omega\big){\rm Im}\Sigma_{+}(\mu-\omega,- \vec p\,)
\nonumber\\
&&-{\rm Im}\Sigma_{+}(\omega-\mu,\vec p\,){\rm Re}
\Sigma_+(\mu-\omega,- \vec p\,)
-{\rm Re}\Sigma_+(\omega-\mu,\vec p\,){\rm Im}
\Sigma_+(\mu-\omega,-\vec p\,).
\end{eqnarray}


\begin{thebibliography}{99}

\bibitem{Klimov:1981ka}
  V.~V.~Klimov,
  Sov.\ J.\ Nucl.\ Phys.\  {\bf 33} (1981) 934
  [Yad.\ Fiz.\  {\bf 33} (1981) 1734];
  Sov.\ Phys.\ JETP {\bf 55} (1982) 199
  [Zh.\ Eksp.\ Teor.\ Fiz.\  {\bf 82} (1982) 336].

\bibitem{Weldon:1982bn}
  H.~A.~Weldon,
  Phys.\ Rev.\ D {\bf 26} (1982) 2789.

\bibitem{Weldon:1989ys}
  H.~A.~Weldon,
  Phys.\ Rev.\ D {\bf 40} (1989) 2410;
  Phys.\ Rev.\ D {\bf 61} (2000) 036003.

  
\bibitem{Pisarski:1989wb}
  R.~D.~Pisarski,
  Nucl.\ Phys.\ A {\bf 498} (1989) 423C.

\bibitem{Braaten:1990it}
  E.~Braaten and R.~D.~Pisarski,
  Phys.\ Rev.\ D {\bf 42} (1990) 2156;
  Nucl.\ Phys.\ B {\bf 337} (1990) 569;
  Phys.\ Rev.\ Lett.\  {\bf 64} (1990) 1338;
  Phys.\ Rev.\ D {\bf 46} (1992) 1829;
  Phys.\ Rev.\ D {\bf 45} (1992) 1827.

\bibitem{Braaten:1991dd}
  E.~Braaten and T.~C.~Yuan,
  Phys.\ Rev.\ Lett.\  {\bf 66} (1991) 2183.

\bibitem{Braaten}
  E.~Braaten,
  Astrophys.\ J. {\bf 392} (1992) 70.
  
\bibitem{Baym:1992eu}
  G.~Baym, J.~P.~Blaizot, and B.~Svetitsky,
  Phys.\ Rev.\ D {\bf 46} (1992) 4043.


\bibitem{Blaizot:1993bb}
  J.~P.~Blaizot and J.~Y.~Ollitrault,
  Phys.\ Rev.\ D {\bf 48} (1993) 1390.

\bibitem{Pisarski:1993rf}
  R.~D.~Pisarski,
  Phys.\ Rev.\ D {\bf 47} (1993) 5589.

\bibitem{Blaizot:1996pu}
  J.~P.~Blaizot,
  Nucl.\ Phys.\ A {\bf 606} (1996) 347.

\bibitem{Thoma:1997dk}
  M.~H.~Thoma and C.~T.~Traxler,
  Phys.\ Rev.\ D {\bf 56} (1997) 198.

\bibitem{Schaefer:1998wd}
  A.~Schaefer and M.~H.~Thoma,
  Phys.\ Lett.\ B {\bf 451} (1999) 195.

\bibitem{Peshier:1998dy}
  A.~Peshier, K.~Schertler, and M.~H.~Thoma,
  Annals Phys.\  {\bf 266} (1998) 162.
  
\bibitem{Peshier:1999dt}
  A.~Peshier and M.~H.~Thoma,
  Phys.\ Rev.\ Lett.\  {\bf 84} (2000) 841.

\bibitem{Mustafa:2002pb}
  M.~G.~Mustafa and M.~H.~Thoma,
  Pramana {\bf 60} (2003) 711.
  
\bibitem{Wang:2004tg}
  S.~Y.~Wang,
  Phys.\ Rev.\ D {\bf 70} (2004) 065011.

\bibitem{Kitazawa:2005pp}
  M.~Kitazawa, T.~Kunihiro, and Y.~Nemoto,
  Phys.\ Lett.\ B {\bf 631} (2005) 157.
  
\bibitem{Bellac}
  M.~Le~Bellac, {\it Thermal Field Theory}
  (Cambridge University Press, Cambridge, 2000).

\bibitem{Review}
  D.~Bailin and A.~Love,
  Phys.\ Rept.\  {\bf 107} (1984) 325;
  M.~G.~Alford, K.~Rajagopal, and F.~Wilczek,
  Phys.\ Lett.\ B {\bf 422} (1998) 247;
  R.~Rapp, T.~Sch\"afer, E.~V.~Shuryak, and M.~Velkovsky,
  Phys.\ Rev.\ Lett.\  {\bf 81} (1998) 53;
  K.~Rajagopal and F.~Wilczek,
  arXiv:hep-ph/0011333;
  M.~G.~Alford,
  Ann.\ Rev.\ Nucl.\ Part.\ Sci.\  {\bf 51} (2001) 131;
  
\bibitem{Rischke:2003mt}
  D.~H.~Rischke,
  Prog.\ Part.\ Nucl.\ Phys.\  {\bf 52} (2004) 197.


\bibitem{Kapusta} 
  J.~I.~Kapusta, 
  {\it  Finite-temperature field theory}
  (Cambridge University Press, Cambridge, 1993).

\bibitem{Fetter+Walecka} 
  A.~L.~Fetter and J.~D.~Walecka, 
  {\it Quantum Theory of many-particle systems}
  (McGraw-Hill Book Company, New York, 1971).
  
\bibitem{Pisarski:1999tv}
  R.~D.~Pisarski and D.~H.~Rischke,
  Phys.\ Rev.\ D {\bf 61} (2000) 074017.

\bibitem{Pisarski:1999av}
  R.~D.~Pisarski and D.~H.~Rischke,
  Phys.\ Rev.\ D {\bf 60} (1999) 094013.


\bibitem{Fugleberg:2002rk}
  T.~D.~Fugleberg,
  Phys.\ Rev.\ D {\bf 67} (2003) 034013.
  
\bibitem{Nickel:2006mm}
  D.~Nickel,
  [arXiv:hep-ph/0607224].

\end{thebibliography}
\end{document}